\documentstyle[epsfig]{mn}
\DeclareMathVersion{bold}
\title[Scalar Statistics]{Scalar statistics on the sphere: application
to the CMB} 
\author[Monteser\'\i n et al.]{C. Monteser\'\i n$^{1,2}$,
R.B. Barreiro$^1$, J.L. Sanz$^1$ and E. Mart\'\i nez-Gonz\'alez$^1$ \\ 
$^1$ Instituto de F\'\i sica de Cantabria, CSIC-Univ. de Cantabria,
Avda. de los Castros s/n, 39005 Santander, Spain\\
$^2$ Dpto. de F\'\i sica Moderna, Univ. de Cantabria, Avda. de
los Castros s/n, 39005 Santander, Spain \\
}
\date{Accepted  Received ; in original form }

\begin{document}
\maketitle

\begin{abstract}
A method to compute several scalar quantities of Cosmic Microwave Background 
maps on the sphere is presented. We consider here four type 
of scalars: the Hessian matrix scalars, the distortion scalars, the gradient 
related scalars and the curvature scalars. Such quantities are obtained 
directly from the spherical harmonic coefficients $a_{\ell m}$ of the map. 
We also study the probability density function of these quantities for
the case of a homogeneous and isotropic Gaussian field, which are
functions of the power spectrum of the initial field.
From these scalars it is posible to construct a new set of scalars which are
independent of the power spectrum of the field.
We test our results using simulations and find a good
agreement between the theoretical probability density functions and
those obtained from simulations. Therefore, these quantities 
are proposed to investigate the presence of non-Gaussian features in CMB
maps. Finally, we show how to compute the scalars in presence of anisotropic 
noise and realistic masks.
\end{abstract}


\section{Introduction}
The Cosmic Microwave Background (CMB) is currently one of the most
valuable tools of cosmology, providing with a great wealth of
information about the universe. A particularly interesting subject is
whether the CMB temperature fluctuations follow or not a Gaussian
distribution. This is a key issue since Gaussian fluctuations are
predicted by the standard inflationary theory, whereas alternative
theories produce non-Gaussian signatures in the CMB. In
addition, foregrounds and systematics may also introduce
non-Gaussianity, which should be carefully studied in order to avoid
its misidentification with intrinsic non-Gaussian fluctuations.

A large effort has been recently devoted to the study of the Gaussian
character of the CMB using the multi-frequency all-sky CMB data
provided by the WMAP satellite of NASA (Bennett et al. 2003a), finding, 
in some cases, unexpected results. 
Some authors have found that the WMAP data are consistent with
Gaussianity using different types of analysis (Komatsu et al. 2003,
Colley \& Gott 2003, Gazta\~naga \& Wagg 2003, Gazta\~naga et
al. 2003).
However, other studies have produced a positive detection of non-Gaussianity 
and/or have shown north-south asymmetries (Chiang et al. 2003,
Eriksen et al. 2004a,b, Park 2004, Copi et al. 2004, Vielva
et al. 2004, Hansen et al. 2004, Mukherjee \& Wang 2004, Land \& Magueijo 
2004, Hansen et al. 2004, Larson \& Wandelt 2004, 
McEwen et al. 2004, Cruz et al. 2005). Although some of these results could be 
explained by the presence of foreground contamination, in other cases 
the origin of the detection is uncertain and a primordial origin can
not be discarded. These results motivates, even more, the development of 
novel techniques to perform further Gaussianity analyses of the CMB. 

In this paper we have focused on the study of several scalar quantities
constructed from the derivatives of the CMB field on the sphere. In
particular, we have considered the modulus of the gradient, the
Laplacian, the distortion, the shear, the ellipticity, the shape index, the
eigenvalues of the negative Hessian matrix, the Gaussian curvature
and the derivative of the squared modulus of the gradient. A procedure to 
calculate the scalars directly from the $a_{\ell m}$ harmonic coefficients is 
provided. In addition, we have obtained expressions for the
probability distributions  
of the previous quantities for the case of a homogeneous and isotropic 
Gaussian random field (HIGRF), which depend on the power spectrum of the
initial field. In order to remove this dependence, we have also
constructed a new set of {\it normalized scalars} whose probability
distribution is independent of the power spectrum.

A number of works have been already devoted to the study of some of
these or other related quantities. Also, some of them have been
applied to study the Gaussianity of the WMAP data, providing
interesting results. In a pioneering work, Bond \& Efstathiou (1987)
studied different statistical properties of the CMB, assuming a
2-dimensional Gaussian field, which included the number densities of
hot and cold spots, the eccentricities of peaks and peak correlation
properties. Barreiro et al. (1997) studied also the mean number of
maxima and the probability distribution of the Gaussian curvature and
the eccentricity of peaks of the CMB for different shapes of the power
spectra. Wandelt, Hivon \& G\'orski (1998) developed efficient
algorithms for fast extrema search as well as for the simulation of
the gradient vector and the curvature tensor fields on the sphere
associated to the temperature field. The power to detect
non-Gaussianity in the CMB of the number, eccentricity and
Gaussian curvature of excursion sets above (and below) a threshold was
tested by Barreiro, Mart\'\i nez-Gonz\'alez \& Sanz (2001) using
Gaussian and non-Gaussian simulations, finding that the Gaussian
curvature was the best discriminator. Dor\'e, Colombi \& Bouchet
(2003) tested the power of a 
technique based on the proportion of hill, lake and saddle points
(which are defined attending to their local curvature) on flat patches
of the sky. A study of the ellipticity of the CMB was performed for the
COBE-DMR (Gurzadyan \& Torres 1997) and Boomerang data
(Gurzadyan et al. 2003), which seemed to indicate a slight 
excess of ellipticity for the largest spots with respect to what was
expected in the standard model. These results were
recently confirmed by repeating the analysis on the WMAP data using the
same region of the sky observed by Boomerang (Gurzadyan et al. 2004).
Eriksen et al. (2004b) applied the Minkowski functionals and the length
of the skeleton (Novikov, Colombi \& Dor\'e 2003) to the WMAP data finding
evidence of non-Gaussianity and asymmetry between the northern and southern
hemispheres. The length of the skeleton is a quantity defined in terms of the
first and second derivatives of the field and the previous work
provided with an algorithm to calculate it on the sphere for a CMB map
in HEALPix pixelization. Hansen et al. (2004) found north/south
asymmetries on the WMAP data using the local curvature, which were
consistent with the results of Eriksen et al. (2004b).
Finally, Cabella et al. (2004) used a method based on the local
curvature to constrain the value of the non-linear coupling constant
$f_{\rm NL}$ from the WMAP data.

The paper is organised as follows. In $\S$2 we introduce the different
considered scalars and describe how to calculate them
from the covariant derivatives of a 2D field and, in particular, for
spherical coordinates. In $\S$3 we obtain analytical (or
semi-analytical) expressions for the probability density
function (pdf) of the considered scalars for a homogeneous and
isotropic Gaussian field on the sphere.
In $\S$4 we construct a new set of related scalars which are independent of 
the power spectrum of the field and we obtain their theoretical distribution 
functions for the Gaussian case. The theoretical results are also compared 
with CMB simulations. $\S$5 shows how to deal with more realistic simulations, 
including anisotropic noise and a mask that covers the Galactic plane
and the point  
sources. In $\S$6 we present our conclusions and outline future applications 
of this work. Appendix~\ref{ap:derivatives} gives the expressions
necessary to  
calculate the derivatives of the field from the $a_{\ell m}$
coefficients. In Appendix~\ref{ap:sums} we present some
useful results regarding spherical harmonic series. Finally,
Appendix~\ref{ap:norm_scalars} gives some guidelines on how to deduce the 
distribution functions of ordinary and normalised scalars.

\section{Scalars on the sphere}
\label{derivatives_scalars}
Let us consider a 2-dimensional field T$(x_1, x_2)$. From the field
derivatives we can construct quantities that are scalars under a 
change of the coordinate system (i. e. regular general transformation 
$(x_1, x_2) \to (x_1^{\prime}, x_2^{\prime})$: 
$s^{\prime}(x_1^{\prime}, x_2^{\prime}) = s(x_1, x_2)$). 
Regarding 1st derivatives, a single scalar can be constructed in terms 
of the ordinary derivative $T_{,i}$. Regarding 2nd covariant derivatives, 
$T_{;ij}$, of T$(x_1, x_2)$ can 
be expressed as a function of the ordinary derivatives $T_{,ij}$ 
and the Christoffel symbols $\Gamma_{ij}^{k}$ as follows
\begin{equation}
T_{;ij}=T_{,ij} - \Gamma_{ij}^{k} T_{,k} 
\end{equation}
To construct linear scalars we need to contract the indices of these covariant 
tensorial quantities.  

The scalars that depend on second derivatives, although can be
expressed as functions of the covariant derivatives, are usually 
defined in terms of the eigenvalues $\lambda_{1} ,
\lambda_{2}$, of the negative Hessian matrix $A$ of the field
T$(x_1, x_2)$
\begin{equation}
A = \left(-T_{;ij}\right).
\end{equation}
 That is, $\lambda_1$ and $\lambda_2$ are the
(negative) second derivatives along the two 
principal directions. In the following we will assume $\lambda_{1} 
> \lambda_{2}$. 

Taking into account the values of $\lambda_1$, $\lambda_2$, we can
distinguish among three type of 
points (e. g. Dor\'e  et al. 2003): hill (both eigenvalues are
positive), lake (both are negative) and saddle ($\lambda_1 > 0,
\lambda_2 < 0$). We will also distinguish between saddle 
points with $|\lambda_1| > |\lambda_2|$, that we will call
saddle$_{-}$, and those with $|\lambda_1| < |\lambda_2|$, saddle$_{+}$
(the sign +,- refers to the sign of the Laplacian, see below).

%A scalar is said of first order when it involves additions of
%derivatives, it is called a second order scalar when it involves
%additions of products of two derivatives and so on. 
%In this way
%$\Theta$ is first order scalars, $\sigma^2$ or $detA$ are second 
%order scalars and $D_{g}$ is a third order scalar.

Hereinafter, we will consider spherical coordinates $(\theta,\phi)$, for which
the metric is given by
\begin{equation}
ds^2 = d\theta^2 + \sin^2 \theta ~ d\phi^2 ~,
\label{d2S}
\end{equation}
which gives the following non-zero Christoffel symbols:
\begin{equation}
\Gamma_{\phi\theta}^{\phi} = \Gamma_{\theta\phi}^{\phi}  =
\frac{\cos\theta}{\sin\theta}, ~~~~ 
\Gamma_{\phi\phi}^{\theta} = - \sin\theta \cos\theta ~.
\label{christoffel_simbols}
\end{equation}

It is convenient to define the following quantities for the spherical
case: 
\begin{equation}
\label{eq:xtheta}
p = \left[T^{,\theta} T_{,\theta}\right]^{\frac{1}{2}} = \frac{\partial T}{\partial \theta} ~,
\end{equation}
\begin{equation}
\label{eq:xphi}
q = \left[T^{,\phi} T_{,\phi}\right]^{\frac{1}{2}} =\frac{1}{\sin \theta} \frac{\partial T}{\partial \phi} ~,
\end{equation}
\begin{equation}
r = \left[T_{\ \ \theta}^{;\theta} T_{\ \ \theta}^{;\theta}\right]^{\frac{1}{2}} =
  \frac{\partial^2 T}{\partial \theta^2} ~,
\end{equation}
\begin{equation}
s = \left[T_{\ \ \phi}^{;\phi} T_{\ \ \phi}^{;\phi}\right]^{\frac{1}{2}}
  =\frac{1}{\sin^2 \theta} \frac{\partial^2 T}{\partial \phi^2} +
  \frac{\cos \theta}{\sin \theta} \frac{\partial T}{\partial \theta} ~, 
\end{equation}
\begin{equation}
\label{eq:x3}
t = \left[T_{\ \ \theta}^{;\phi} T_{\ \ \phi}^{;\theta}\right]^{\frac{1}{2}} =
  \frac{1}{\sin \theta} \frac{\partial^2 T}{\partial \theta \partial
  \phi} - \frac{\cos \theta}{\sin^2 \theta} \frac{\partial T}{\partial
  \phi} ~.
\end{equation}
We will express all the scalars as a function of these five quantities. 
All of them, and therefore the scalars, can be easily computed from
the $a_{\ell m}$ coefficients of the field (see
Appendix~\ref{ap:derivatives}). 

In the following, we present some of these scalars as 
function of the covariant derivatives. We will also relate them to the former
defined quantities for the spherical case.

Note that, although we are considering the spherical case, the
expressions given for the scalars as a function of the covariant
derivatives are valid for any 2-dimensional surface. In particular,
if observing a small portion of the sky, we can assume that we have a
flat patch. In this case, the scalars can be easily obtained taking
into account that all the Christoffel symbols are zero.

\subsection{The Hessian matrix scalars}
We include in this section the algebraical quantities related to the 
Hessian matrix: eigenvalues, trace and determinant. 

\subsubsection{The eigenvalues} \label{sec:eigenvalues}

The eigenvalues of the negative Hessian matrix are scalars 
of the field T, and they can be expressed as follows
\begin{equation}
2\lambda_{1}= - \left(T_{\ \ i}^{;i}\right) + 
\sqrt{\left(T_{\ \ i}^{;i}\right)^2 - 2 \left(T_{\ \ i}^{;i}T_{\ \ j}^{;j} - T_{\ \ i}^{;j} T_{\ \ j}^{;i}
\right)} ~, 
\end{equation}
\begin{equation}
2\lambda_{2} = - \left(T_{\ \ i}^{;i}\right) - 
\sqrt{\left(T_{\ \ i}^{;i}\right)^2 - 2 \left(T_{\ \ i}^{;i}T_{\ \ j}^{;j} - T_{\ \
i}^{;j} T_{\ \ j}^{;i} 
\right)} ~.
\end{equation}
For the spherical coordinate system, we can rewrite them in the
following way:
\begin{equation}
2\lambda_{1} = -\left( r + s \right) +
\sqrt{( r - s )^2 + (2 t )^2} ~,
\label{eq:lambda1}
\end{equation}
\begin{equation}
2\lambda_{2} = -\left( r + s \right) -
\sqrt{\left( r - s \right)^2 + \left(2 t \right)^2} ~. 
\label{eq:lambda2}
\end{equation}
Note that $\lambda_{1} \geq \lambda_{2}$. Therefore for values of
$\lambda_{2}>0$ we have hill points whereas values of $\lambda_{1}<0$
correspond to lake points. 
These eigenvalues coincide with the two principal curvatures of the surface at
extrema points where the first derivatives are zero. 

\subsubsection{The Laplacian} \label{sec:Laplacian}

The Laplacian $\lambda_{+}$ is defined as the trace of the 
 Hessian matrix. Therefore it can be expressed as a function of 
the eigenvalues: 
\begin{equation}
\lambda_{+} = -\lambda_{1} - \lambda_{2} ~.
\end{equation} 
Note that negative values of $\lambda_{+}$ correspond to hill or saddle$_{-}$
points of the field, whereas lake and saddle$_{+}$ points correspond to
positive values of the Laplacian.

The Laplacian on the sphere can also be written as a function of the 
field covariant derivatives or $r$ and $s$:

\begin{equation}
\lambda_{+} = T_{ i}^{;i}= r + s  ~.
\end{equation}
%Note that at the extrema the Laplacan coincides with the intrinsic
%curvature $k = \frac{1}{2} \frac{T^{;ij} T_{i}
%T_{j}-T_{i}^{;i}}{\sqrt{1 +T^{i} T_{i}}}$. 

\subsubsection{The determinant of A} \label{sec:Det_A}

Another scalar that can be constructed is the determinant of the negative
Hessian matrix $A$, which is given by
\begin{equation}
d \equiv det A = \lambda_{1} \lambda_{2} ~.
\end{equation} 
Positive values of $d$ correspond to hill or lake points of the
field whereas saddle points are given by negative values of $d$.

As a function of the covariant derivatives, the determinant can be
expressed as
\begin{equation}
d = \frac{1}{2}\left[T_{\ \ i}^{;i} T_{\ \ j}^{;j} - T_{\ \ i}^{;j}
T_{\ \ j}^{;i}\right] ~.
\end{equation}
%And it coincides with the Gaussian curvature $\kappa_{G}$ in maxima and minima.
%
Finally, for the spherical coordinate system, we can rewrite it as
\begin{equation}
d = r ~ s  - t^2 ~.
\label{eq:determinant}
\end{equation}

\subsection{The distortion scalars}
We consider here the distortion, the shear, the ellipticity and the shape 
index. They are all related with powers of the difference of the eigenvalues 
$\lambda_{1}-\lambda_{2}$, so they give us information about the distortion
of the field.
 
\subsubsection{The shear} \label{sec:shear}
An interesting scalar related to the eigenvalues is the shear,
which is defined as
\begin{equation}
y =\frac{1}{4} (\lambda_{1}-\lambda_{2})^2 ~. 
\end{equation}
As a function of the covariant derivatives we can express it as follows
\begin{equation}
y = \frac{1}{4} \left[T_{\ \ i}^{;i}\right]^2 -\frac{1}{2}\left[T_{\ \ i}^{;i}
 T_{\ \ j}^{;j} - T_{\ \ i}^{;j} T_{\ \ j}^{;i}\right] ~, 
\end{equation}
and for the spherical coordinate system, we can rewrite it as
\begin{equation}
y =\frac{1}{4}( r - s )^2 + t^2 ~. 
\end{equation}

\subsubsection{The distortion} \label{sec:distortion}
The distortion is defined as the difference of the eigenvalues of the negative 
Hessian matrix:
\begin{equation}
\lambda_{-} = \lambda_{1}-\lambda_{2} 
\end{equation}
therefore, by construction $0 < \lambda_{-} < \infty$. We can express the 
distortion as a function of the covariant derivatives of the original field
in the following manner
\begin{equation}
\lambda_{-} = \sqrt{ \left[T_{\ \ i}^{;i}\right]^2 - 2 \left[T_{\ \ i}^{;i}
 T_{\ \ j}^{;j} - T_{\ \ i}^{;j} T_{\ \ j}^{;i}\right]}
\end{equation}
using the spherical coordinate system, we can rewrite it as
\begin{equation}
\lambda_{-} =\sqrt{\left(r-s \right)^2 + \left(2t \right)^2}. 
\end{equation}

\subsubsection{The ellipticity} \label{sec:Ellipticity}

The ellipticity is defined as
\begin{equation}
e = \frac{\lambda_{1} - \lambda_{2}}{2(\lambda_{1} + \lambda_{2})} ~.
\end{equation} 
It can take values in the whole real domain. It is
straightforward to show that $e \in \left(-\frac{1}{2},0\right)$ and $e \in \left(0 , \frac{1}{2}\right)$ for 
lake and hill points respectively. Regarding to saddle points,
the ellipticity takes values in the range $e \in \left(\frac{1}{2} , \infty \right)$ for $saddle_{-}$
and $e \in \left(-\infty , -\frac{1}{2} \right)$ for $saddle_{+}$.

As a function of the covariant derivatives of the field we can express 
the ellipticity in the following way:
\begin{equation}
e = - \frac{1}{2 \left[T_{\ \ i}^{;i}\right]}
\sqrt{\left[T_{\ \ i}^{;i}\right]^2 - 2 \left[T_{\ \ i}^{;i}T_{\ \
j}^{;j} - T_{\ \ i}^{;j} 
 T_{\ \ j}^{;i} \right]} ~,
\end{equation} 
and using $r$, $s$ and $t$ in spherical coordinates the ellipticity
 is rewritten as follows:
\begin{equation}
e = -\frac{\sqrt{( r - s )^2 + (2t)^2}}{2( r + s )} ~.
\label{eq:ellipticity}
\end{equation} 

\subsubsection{The shape Index}

There are also other scalars that can be constructed from the
ellipticity. One of the most interesting quantities is the shape index 
(Koenderink 1990), defined as
\begin{equation}
\iota=\frac{2}{\pi} \arctan{\left(-\frac{1}{2e}\right)} ~.
\label{eq:shape_index_def}
\end{equation}
By definition, the shape index is bound $\iota \in \left(-1 ,1 \right)$.
 Hill points correspond to 
values of $\iota \in \left(-1 ,-\frac{1}{2} \right)$ , lake points to 
$\iota \in \left(\frac{1}{2} , 1 \right)$, saddle$_{-}$ points to
 $\iota \in \left(-\frac{1}{2} , 0 \right)$ and finally 
saddle$_{+}$ points are in the range $\iota \in \left(0 , \frac{1}{2} \right) $.

%Other studied scalar bound construction with the ellipticity is the 
%eccentricity, but its definition depends on the current type of point, 
%for hill and saddle points we use the expression on the left and the 
%one on the right for lake points 
%%
%\begin{equation}
%\epsilon=\sqrt{1-\frac{\lambda_{2}}{\lambda_{1}}} ~~
%\epsilon=\sqrt{1-\frac{\lambda_{1}}{\lambda_{2}}}
%\end{equation}  
%%
%With this definition hill and lake points corresponds to 
%$0<\epsilon<1$, and saddle points to $1<\epsilon<\infty$.

\subsection{The gradient related scalars}

We included here the squared modulus of the gradient field, and 
also another quantity related with its derivative.

\subsubsection{The square of the gradient modulus} 
\label{sec:gradcuad}

The square of the modulus of the gradient is defined as follows:
\begin{equation}
g \equiv \mid \vec{\nabla} T \mid^2= T^{,i}T_{,i} ~.
\label{eq:gradient}
\end{equation}
It gives information about the smoothness of the field. Taking into
account equations~(\ref{eq:xtheta}) and~(\ref{eq:xphi}), 
$g$ can be expressed for the spherical coordinate system as
\begin{equation}
g = p^2 + q^2 ~. 
\end{equation}

\subsubsection{Derivative of the square of gradient modulus} \label{sec:Dg} 

We also consider another non linear scalar, called $D_{g}$, which is 
proportional to the derivative of the square of the gradient modulus
 with respect to the arc associated to the integral curves of
 the gradient. As a function of the covariant derivatives we can 
write it in the following way: 
\begin{equation}
D_{g} = \frac{d}{d s} \left(\frac{1}{2} g \right) = T^{;ij}T_{,i}T_{,j} ~.
\label{eq:Dg}
\end{equation}
In the spherical coordinate system we can express it as
\begin{equation}
D_{g} = r ~ p^2 + 2 ~ t ~ p ~ q + s ~ q^2 ~,
\end{equation}  
where $D_{g} \in \left(-\infty , \infty \right)$. 
This quantity is related to the extrinsic curvature (see below).

\subsection{The curvature Scalars}
We include in this section the intrinsic and extrinsic curvatures. They
give us information about the geometry of the surface of the temperature
field. 
 
\subsubsection{The Gaussian curvature} \label{sec:gaussian_curvature}

The Gaussian curvature, also called intrinsic curvature, is defined as the
product of the two principal curvatures of the surface defined by the 
temperature field. It can be expressed as a function of the field derivatives: 
\begin{equation}
\kappa_{G} = \frac{1}{2}\frac{T_{\ \ i}^{;i} T_{\ \ j}^{;j} - T_{\ \
i}^{;j} T_{\ \ j}^{;i}}{\left[1+\left(T^{,i}T_{,i}\right)\right]^2} ~, 
\label{eq:gaussian_curv}
\end{equation}
 we can rewrite it in the spherical coordinate system as follows:
\begin{equation}
\kappa_{G} = \frac{ r ~ s - t^2}{\left[1+ p^2 + q^2\right]^2} ~,
\label{eq:gaussian_curv2}
\end{equation} 
where $\kappa_{G} \in \left(-\infty , \infty \right)$ by construction.
Note that for extrema, the Gaussian curvature coincides with the
determinant of $A$. This scalar depends on intrinsic 
properties of the surface.

\subsubsection{The extrinsic curvature}
This scalar is defined as the average of the two principal curvatures of
the field and it can be rewritten as a function of previous scalars in the 
following way: 
\begin{equation}
\kappa_{ex}=\frac{1}{2} ~ \frac{1}{\sqrt{1+g}}\left[\lambda_{+} -
\frac{D_{g}}{\left(1+g\right)}\right] ~,
\label{eq:extrinsic_curvature}
\end{equation}
This scalar gives an idea of how the surface generated by the temperature field is embedded in $\Re^{3}$.
Note that in the extrema this quantity coincides with the Laplacian except by a
constant factor.

\section{Homogeneous and isotropic Gaussian random field on the sphere}
\label{gaussian_case}
In this section, we will derive the pdf of the scalars defined in the preceding
section for a HIGRF on the sphere.
Under this assumption, it can be shown that the quantities $\{p, q, r, s, t\}$ 
defined in equations (\ref{eq:xtheta}) to (\ref{eq:x3}), follow also a homogeneous
 and isotropic Gaussian distribution. In order to calculate the dispersions
 and covariances of these quantities, we define the moments $\sigma_i$ as
\begin{equation}
\sigma_{i}^2 = \sum_{\ell} C_{\ell}
\frac{2\ell+1}{4\pi}\left[\ell\left(\ell+1\right)\right]^{i} ~,
\label{eq:momentos}
\end{equation}
where $C_{\ell}$ is the power spectrum of the temperature field.
Making use of the results given in Appendix~\ref{ap:sums}, it can be
shown that:
\begin{eqnarray}
<p ~ p> & = &<q ~ q> ~ = ~ \frac{1}{2}\sigma_{1}^2 ~,\\
<r ~ r> & = &<s ~ s> ~ = ~ \frac{3}{8}\sigma_{2}^{2} -
\frac{1}{4}\sigma_{1}^{2} ~,\\ 
<t ~ t> & = &\frac{1}{8}\sigma_{2}^{2} - \frac{1}{4}\sigma_{1}^{2} ~.
\end{eqnarray}
If the initial field has zero mean, $\{p, q, r, s, t\}$ have also zero mean, 
and their covariances are zero except for the term $< r ~ s>$ which is 
given by:
\begin{equation}
<r ~  s>=\frac{1}{8}\sigma_{2}^{2} + \frac{1}{4}\sigma_{1}^{2} \equiv
\rho <r ~ r> \, .
\end{equation}
In order to obtain the pdf's of the scalars, it is useful to define a new
set of Gaussian variables which are uncorrelated among them. In
particular, we construct $R = r + s$, $S = r - s$ and 
$T = 2 t$. It is straightforward to show that
$\sigma_{R}^{2}=\sigma_{2}^{2}$ and that $\sigma_{S}^{2} =
\sigma_{T}^{2} = \frac{1}{2} \sigma_{2}^{2} - \sigma_{1}^{2}$.  

As an illustration, in Fig.~\ref{fig:x1,xphi} we show the theoretical
distribution functions of 
the variables $q$ and $r$ as well as the ones obtained
averaging the normalised histograms of 20 all-sky CMB Gaussian simulations.
The simulations have been generated using the HEALPix package
(G\'orski et al. 1999) for $N_{\rm side}=256$ and using the power 
spectrum given by the 
best-fit model found by the WMAP team (Spergel et al. 2003). The $C_\ell$'s
for this model were generated using CMBFast (Seljak \& Zaldarriaga
1996). The simulated maps were smoothed with a Gaussian beam of full
width half maximum (FWHM) equal to 2.4 times the pixel size. Note the good
agreement between the theoretical distribution and the numerical results.
\begin{figure}
\begin{center}
\includegraphics[angle=0, width=7.0cm]{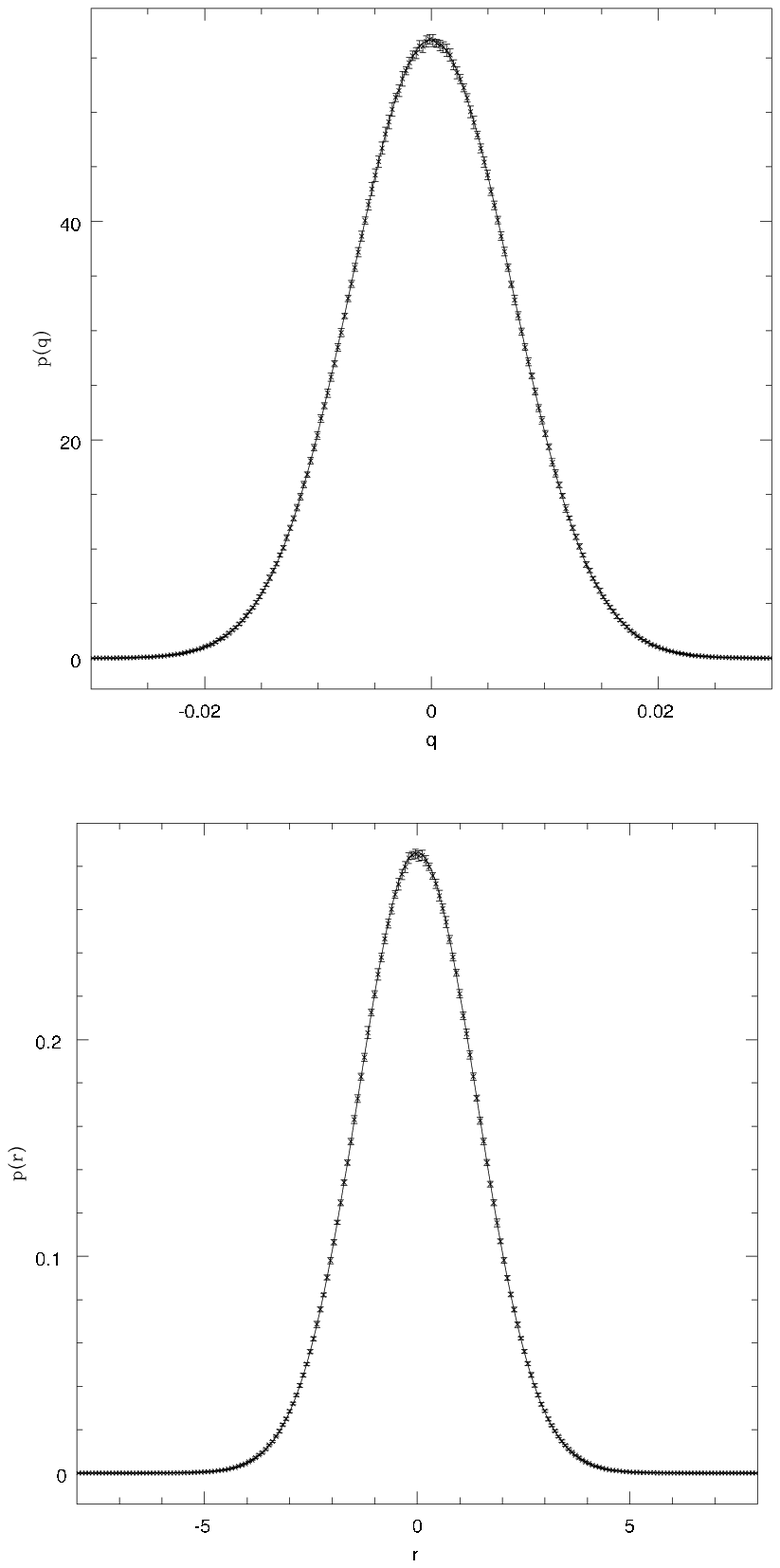}
\caption{Top panel: the solid line corresponds to the theoretical
distribution expected for $q$, the crosses have been obtained
averaging over 20 all-sky CMB simulations and the error bars give the
dispersion for the same 20 simulations. Bottom panel: the same results
are shown for the variable $r$.}
\label{fig:x1,xphi}
\end{center} 
\end{figure}

Since we have expressed all the scalars as a function of
$\{p, q, r, s, t\}$ and since they follow a simple
Gaussian distribution, we can almost straightforwardly derive the
probability distribution function of the considered scalars.

%To test the theoretical distributions, they will be compared with the
%ones obtained from 20 full-sky CMB Gaussian simulations, generated as
%explained before. A map of the studied scalar is then obtained for each
%simulation and a normalised histogram of the scalar is
%constructed. Finally, the average value and dispersion of the histogram from
%the 20 simulations is computed and plotted versus the theoretical
%distribution.

\subsection{The Hessian matrix scalars}

\subsubsection{The eigenvalues} \label{sec:eigenvalues_pdf}
Taking into account equation (\ref{eq:lambda1}), we can obtain the pdf
of $\lambda_1$:

\begin{equation}
p\left(\lambda_1\right) = 
f e^{-2\frac{\lambda_1 ^2}{\sigma_{2}^2}} + k \lambda_1 e^{-c
\lambda_1^2}  
\left[1 + {\rm erf}\left(h \lambda_1\right)\right] ~,
\label{eq:pdf_lambda_1}
\end{equation}
where the constants $c, f, k$ and $h$ are given by
\begin{eqnarray}
c & = &\frac{2}{\sigma_{S}^2 +\sigma_{2}^2 } ~~,~~
f ~ = ~\frac{\sigma_{2}}{\sqrt{2\pi}} c \nonumber ~,\\
k & = &\frac{\sigma_{S}}{\sqrt{2}} c^{\frac{3}{2}}    ~~~~~,~~
h ~ = ~\frac{\sigma_{S}}{\sigma_{2}} c^{\frac{1}{2}} ~,
\end{eqnarray}
and erf is the error function.
For this distribution function we obtain $<\lambda_{1}>=
\sqrt{\frac{\pi}{8}} \sigma_{S}$.

Analogously, we can obtain the pdf for $\lambda_{2}$ :
\begin{equation}
p\left(\lambda_2\right) = 
f e^{-2\frac{\lambda_2 ^2}{\sigma_{2}^2}} - k \lambda_2 e^{-c
\lambda_2^2}  
\left[1 - {\rm erf}\left(h \lambda_2\right)\right] ~. 
\label{eq:pdf_lambda_2}
\end{equation}
In this case, we find $<\lambda_{2}>=-\sqrt{\frac{\pi}{8}} 
\sigma_{S}$. 
Note that there is a symmetry between the pdf's of $\lambda_1$
and $-\lambda_2$, such that $p_{\lambda_1}(x)=p_{\lambda_2}(-x)$.

\begin{figure}
\begin{center}
\includegraphics[width=\hsize]{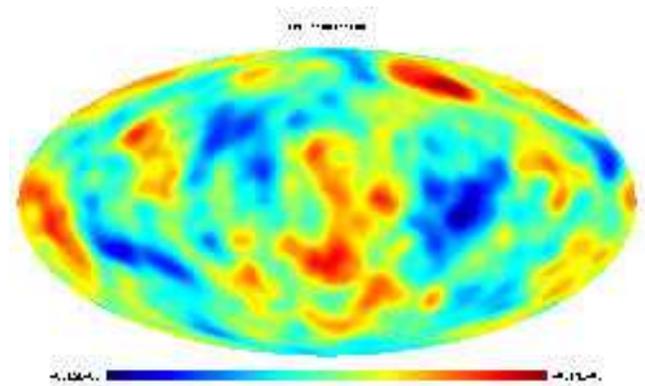}
\caption{A Gaussian CMB simulation (in units of $K$) smoothed with a 
Gaussian beam of FWHM=$8^\circ$ is shown.}
\label{fig:map_temp}
\end{center}
\end{figure}

The map of the eigenvalues $\lambda_1$ and $\lambda_2$, corresponding to the 
CMB Gaussian simulation in Fig.~\ref{fig:map_temp} (smoothed with a Gaussian 
beam of FWHM of 8 deg) can be seen in the top
left and right panel of Fig.~\ref{fig:map_loc} respectively.
 We have chosen a low resolution
CMB map for a better visualization of the different type of points, that with
this smoothing appears connected, since otherwise the maps of the
scalars are dominated by the small scale and the large scale structure
can not be appreciated. 
Note that negative values of 
$\lambda_{1}$, corresponding to lake points in the original field, form 
compact regions while the rest of pixels, corresponding to other type of
points, form a filamentary structure similar to a web. In a similar way
positive values of $\lambda_{2}$, corresponding to hill points, form compact 
regions surrounded by the filamentary structure.   

\subsubsection{The Laplacian} \label{sec:Laplac_pdf}
\begin{figure*}
\begin{center}
\includegraphics[width=\hsize]{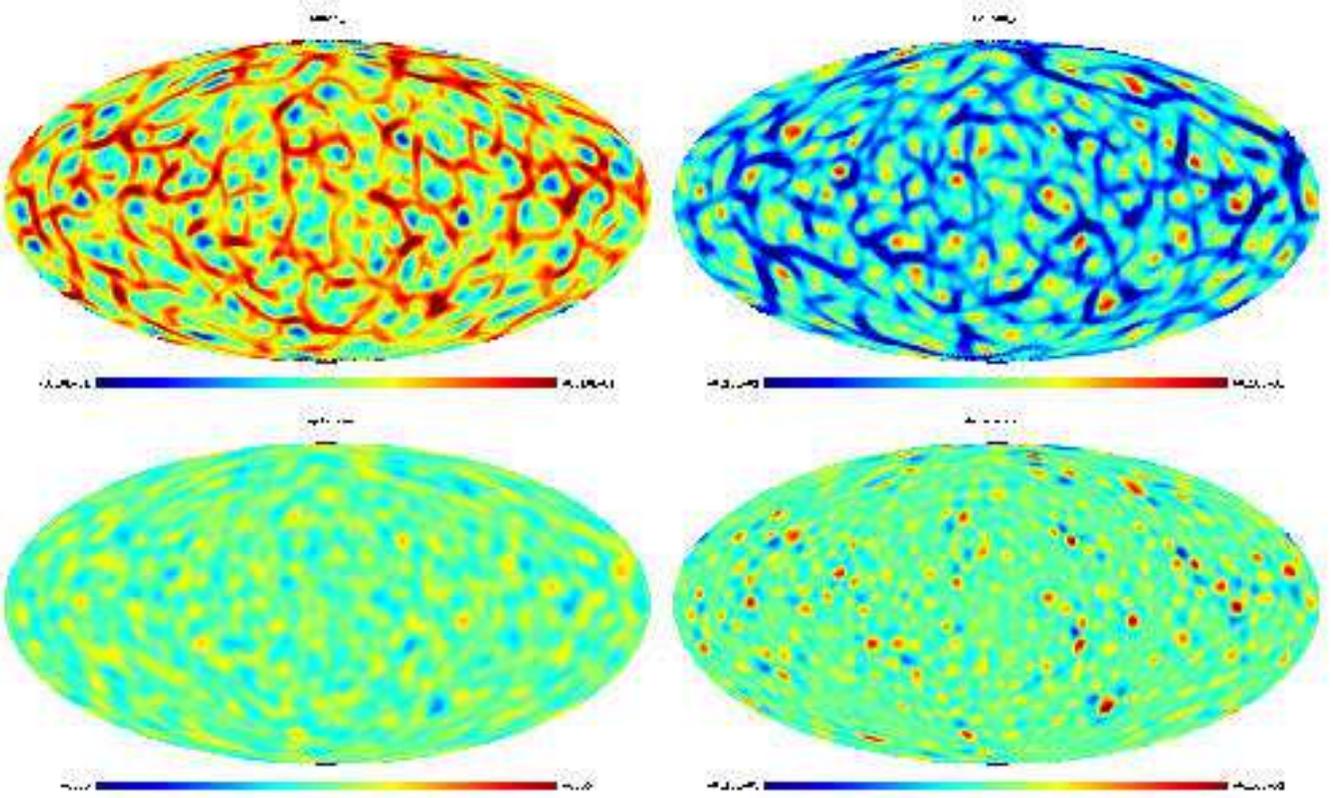}
\caption{Hessian matrix scalars associated to the temperature map of 
Fig.~\ref{fig:map_temp} are shown: the eigenvalues $\lambda_1$ (top left) and 
$\lambda_2$ (top right), the laplacian (bottom left) and the determinant
 (bottom right).}
\label{fig:map_loc}
\end{center}
\end{figure*}

The Laplacian is the addition of two Gaussian variables with the same 
dispersion and therefore it also follows a Gaussian distribution with 
dispersion $\sigma_{2}$. 

%It is convenient to consider the normalised
%Laplacian $x=-\frac{\tilde_{+}}{\sigma_{2}}$, which simply follows 
%a Gaussian distribution of zero mean and unit dispersion:
%
\begin{equation}
p\left(\lambda_{+}\right) = \frac{1}{\sqrt{2\pi}\sigma_{2}}\exp^{-\frac{1}{2 \sigma_{2}^{2}} \left(\lambda_{+}\right)^{2}} ~.
\end{equation}
The Laplacian is defined in the interval $\lambda_{+} \in \left(-\infty , 
\infty \right)$.
As an example, we show the Laplacian of a CMB
Gaussian simulation in Fig.~\ref{fig:map_loc} (bottom left panel).
Note that those regions with low values of $\lambda_{+}$ correspond to 
regions with high positive curvature in the initial temperature field, and 
analogously regions with high values of $\lambda_{+}$ correspond to regions 
with high negative curvature in $T$.

\subsubsection{The determinant of A} \label{sec:det_A_pdf}
Using equation (\ref{eq:determinant}), we can construct a semi-analytical
expression for the pdf of the determinant: 
\begin{equation} 
p\left(d\right)=C_{d}
e^{wd} \int_{0}^{\infty} e^{a x^2} K_{0}\left(\frac{w}{\rho}
|x^2+d|\right) dx ~,
\end{equation}
where $d \in \left(-\infty , \infty \right)$ and the constants are
\begin{eqnarray}
C_{d} &=&\frac{2}{\pi
\sqrt{2\pi\left(1-\rho^2\right)}}\frac{1}{\sigma_{r}^2 \sigma_{t}}~,
\nonumber \\
w&=&\frac{\rho}{1-\rho^2}\frac{1}{\sigma_{r}^2}~, ~~~ 
a~=~w-\frac{1}{2 \sigma_{t}^2} ~.
\end{eqnarray} 
and $K_0$ is the zero order modified Bessel function of second kind.

A map of $d$ for the Gaussian CMB simulation shown in Fig.~\ref{fig:map_temp}, 
is given in Fig.~\ref{fig:map_loc} (bottom right panel). Positive values of 
$d$ correspond to hill and lake points, so the most curved  hills and 
valleys in the original field are given by the highest values of $d$.

\subsection{The distortion scalars}

\subsubsection{The shear}
\begin{figure*}
\begin{center}
\includegraphics[width=\hsize]{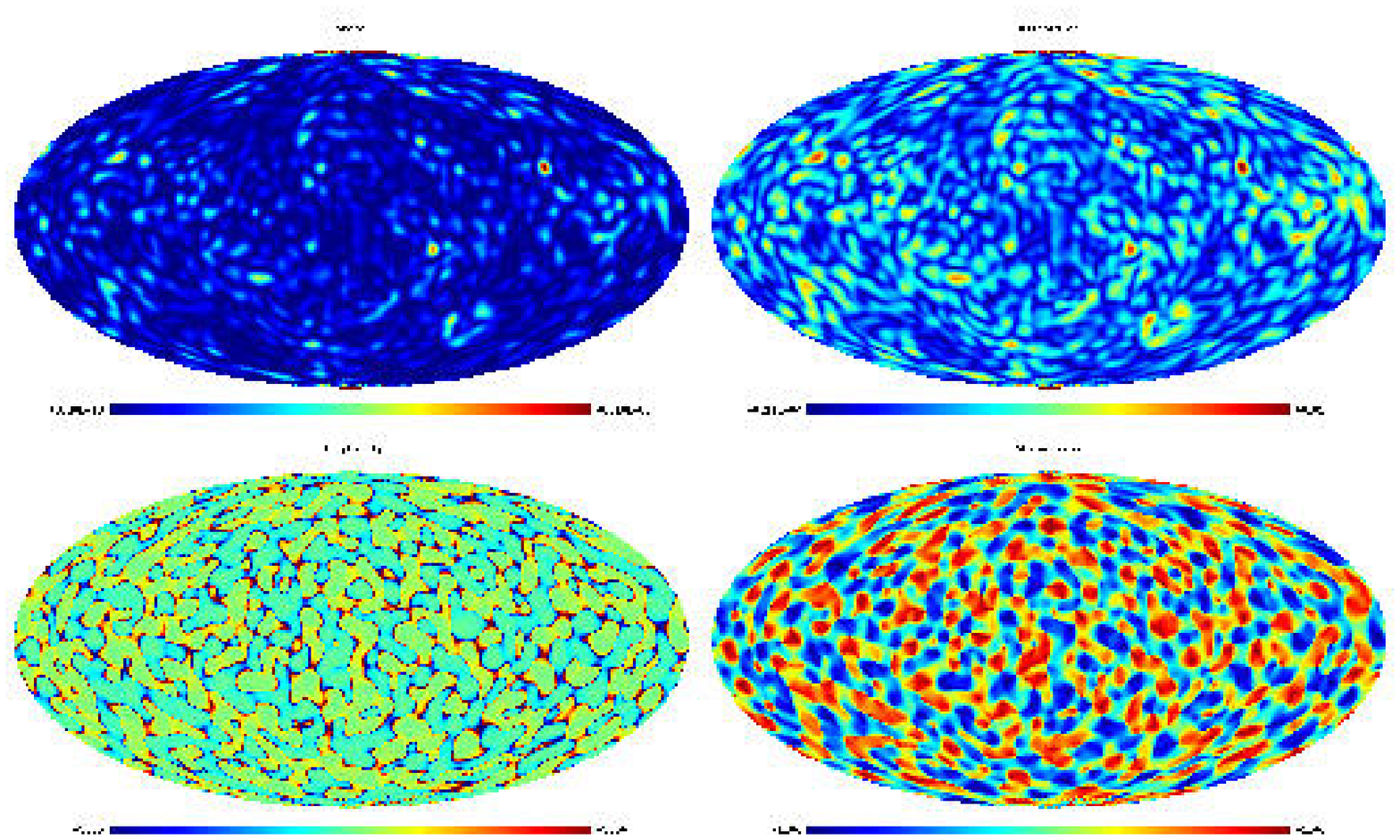}
\caption{The distortion scalars associated to the temperature 
 map presented in Fig.~\ref{fig:map_temp} are plotted. They are the
shear (top left), 
the distortion (top right), the ellipticity (bottom left) and the shape index 
(bottom right).}
\label{fig:map_distor}
\end{center}
\end{figure*}

{\label{sec:Shear_pdf}}
The shear is the addition of two independent squared Gaussian
variables with the same dispersion. Therefore, it follows a
$\chi^2_2$ distribution function with mean and dispersion equal to
$<y>=\sigma_{y}=\frac{1}{4}\sigma_{2}^{2} - \frac{1}{2}\sigma_{1}^{2}$.
\begin{equation}
p\left(y\right)=\frac{1}{\sigma_{y}}e^{-\frac{y}{\sigma_{y}}}  ~,
\end{equation}
where $y \in [0 , \infty) $. The map of the shear
of a Gaussian CMB simulation is shown in the top left panel of
Fig.~\ref{fig:map_distor}. In the considered example, we find that large values
 of $y$ are concentrated in regions which usually correspond to saddle points 
with high values of $\mid \lambda_{i}\mid$ in the original map. 

\subsubsection{The distortion}

The distortion is proportional to the square root of the shear, thus its 
probability density function can be easily obtained from the one of the shear. 
\begin{equation}
p(\lambda_{-}) = \frac{1}{\sigma_{y}} \frac{\lambda_{-}}{2} e^{-\frac{\lambda_{-}^{2}}{4\sigma_{y}}}
\end{equation}
where $\lambda_{-}$ is positive by construction. In the top right panel of 
Fig.~\ref{fig:map_distor} we show the distortion map for the Gaussian CMB 
simulation of Fig.~\ref{fig:map_temp}. The physical information 
contained in the distortion is basically the same as that of the shear. In 
fact, both maps show quite similar structure.

\subsubsection{The ellipticity} \label{sec:Ellipticity_pdf}
Taking into account equation~(\ref{eq:ellipticity}), we can obtain the
pdf of the ellipticity:
\begin{equation}
p\left(e\right)=2\sigma_{S}^{}\sigma_{R}^2
\left|e\right| \left[\frac{1}{\sigma_{S}^2+4\sigma_{R}^2 e^2}
\right]^{\frac{3}{2}} ~,
\label{eq:pdf_ellipticity}
\end{equation}
where $ e \in \left(- \infty , \infty \right)$. The map of
the ellipticity computed for a Gaussian CMB simulation is given in
Fig.~\ref{fig:map_distor} (bottom left panel). It is seen that the largest 
values of $ \mid e \mid$ correspond to saddle points surrounding 
hill or lake points which form compact regions.

\subsubsection{The shape index} \label{sec:shape_index_pdf}
The shape index pdf can be easily derived from the one of the ellipticity:
\begin{equation}
p\left(\iota \right)=\frac{\sigma_{S}\sigma_{R}^2\pi}{4}
\frac{\left|\cos{\left(\frac{\pi}{2}\iota \right)}\right|}
{\left|\sin^3{\left(\frac{\pi}{2}\iota \right)}\right|} 
\left[\sigma_{S}^2 + \sigma_{R}^2
\frac{\cos^2{\left(\frac{\pi}{2}\iota \right)}}{\sin^2{\left(\frac{\pi}{2}\iota \right)}}\right]
^{-\frac{3}{2}} ~.
\end{equation}
Note that the shape index is bound and takes values in the range
$\iota \in \left[-1 , 1 \right]$. The structure of the shape index for a Gaussian 
CMB simulation is shown in the bottom right panel of Fig.~\ref{fig:map_distor}.
 This scalar presents a similar structure as the one of the ellipticity. 
Higher values of $\iota$ correspond to lake points in the original temperature 
field while lower values of $\iota$ correspond to hill points.

\subsection{The gradient related scalars}
    
\subsubsection{The square of the gradient modulus} \label{sec:gradcuad_pdf}
\begin{figure}
\begin{center}
\includegraphics[width=\hsize]{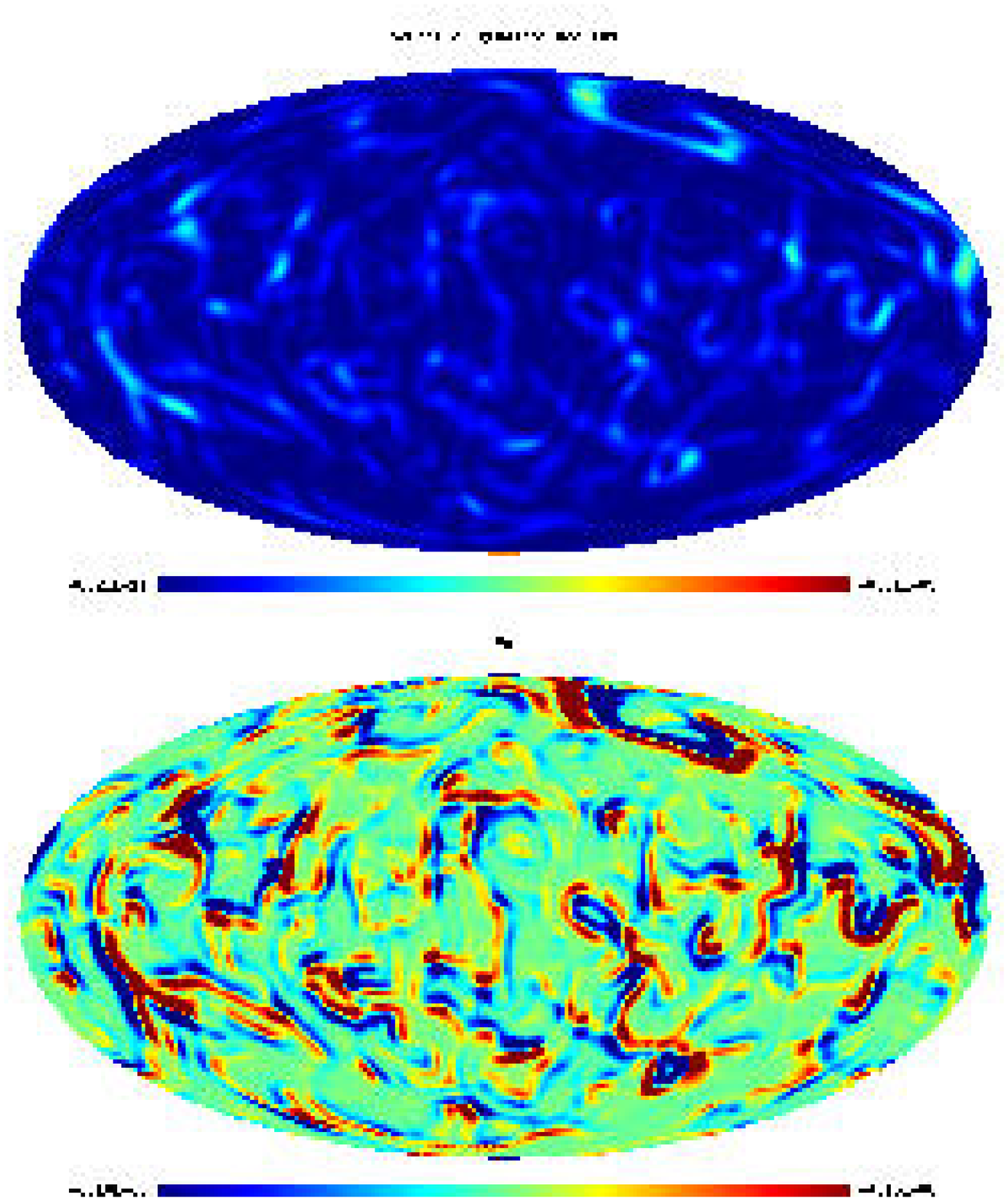}
\caption{Gradient related scalars associated to the temperature map of 
Fig.~\ref{fig:map_temp} are shown: the squared modulus of the gradient
 (top), and the gradient derivative (bottom).}
\label{fig:map_grad}
\end{center}
\end{figure}

Taking into account equation~(\ref{eq:gradient}), we see that $g$ is
given by the addition of two independent squared 
Gaussian variables with the same dispersion. Therefore this scalar 
follows a $\chi^2_2$ distribution with mean and dispersion equal to
$<g>=\sigma_{g}=\sigma_{1}^{2}$ 
\begin{equation}
p(g)=\frac{1}{\sigma_{g}}e^{- \frac{g}{\sigma_{g}}} ~,
\end{equation}
where $g \in [0 , \infty)$. 
As an illustration of the structure of the squared modulus of the
gradient, we show the gradient map of the low resolution CMB Gaussian
 simulation of Fig.~\ref{fig:map_temp} in the top panel of
Fig.~\ref{fig:map_grad}. 
Note that those regions where the temperature field changes rapidly
correspond to high values of $g$. 

\subsubsection{Derivative of the square of gradient modulus} \label{D_g_pdf}
$D_{g}$ involves the five Gaussian variables ~(\ref{eq:xtheta})
to ~(\ref{eq:x3}) and its probability distribution can be expressed as
a function of one integral:
\begin{equation}
p\left(D_{g}\right)= C_{D_{g}} 
\int_{0}^{\infty}\frac{dz}{z}
exp{\left[-\left(\frac{z^2}{2\sigma^{2}_{p}} + 
\frac{D_{g}^2}{2\sigma_{r}^2 z^4}\right)\right]} ~, 
\end{equation}
where $D_{g} \in \left(-\infty , \infty \right)$ and the constant $C_{D_{g}}$ is given by
\begin{equation}
C_{D_{g}}=\frac{1}{\sqrt{2\pi}}
\frac{1}{\sigma_{r}\sigma_{p}^2} ~. 
\end{equation}
 The structure of this scalar is shown in the map of Fig.~\ref{fig:map_grad} 
(bottom panel) obtained for a Gaussian CMB simulation. Note that this scalar 
was defined as the derivative of the squared modulus of the gradient vectorial 
field, except by a constant factor. In this way it gives us extra information 
about the directional variation of the gradient, which is not given by $g$.

\subsection{The curvature scalars}
As shown in equations (\ref{eq:gaussian_curv2}) and
 (\ref{eq:extrinsic_curvature}) the curvature scalars are a function of
 $1+\mid\vec{\nabla} T\mid^2$, which depend on the units of the temperature
 field and, therefore the shape of the curvature scalars pdf's will also 
non-trivially depend on the these units (note that for the previously 
studied scalars, the units of the field enter just as a normalization factor
 in the pdf). For this reason we use adimensional fields to calculate 
the curvatures. In particular, we choose to normalise the initial 
temperature field to unit dispersion to enhance possible deviations from
 Gaussianity.

\subsubsection{Gaussian curvature} \label{sec:gaussian_curvature_pdf}
The pdf of the Gaussian curvature can not be obtained in an analytical
form but it can be written as a function of two integrals:
\begin{equation}
  p \left( \kappa_{G} \right) = C_{\kappa} \int_{0}^{\sqrt {\frac{\rho}{w} }}  dv \frac{1}{v^{4}} 
  e^{\rho \frac{\kappa_{G}}{v^{2}} - b \sqrt{\frac{\rho}{w}} \frac{1}{v}} I\left( \frac{\kappa_{G}}{v^{2}} \right) ~, 
\end{equation}
where $\kappa_{G} \in \left(-\infty , \infty \right)$, and $I \left( \alpha \right)$ is the following integral:
\begin{equation}
I \left( \alpha \right) = \int_{0}^{\infty} dx e^{-\frac{a \rho}{w} x^2}
K_{0}\left(\left|x^2+ \alpha \right|\right) ~. 
\end{equation}
The constants are given by
\begin{eqnarray}
b &=& \frac{1}{2 \sigma_{p}^{2}}~,~
C_{\kappa} ~=~ \frac{e^{b}}{\pi \sqrt{2 \pi} \sqrt{1-\rho^2}}
 \left(\frac{\rho}{w} \right)^2 \frac{1}{\sigma_{r}^2\sigma_{t}\sigma_{p}^{2}} \\
a &=& \frac{1}{2 \sigma_{t}^2} - \gamma ~,~
w ~=~ \frac{\rho}{1-\rho^2} \frac{1}{\sigma_{r}^2}. 
\end{eqnarray}
We present in the top panel of Fig.~\ref{fig:map_curv} the Gaussian
curvature corresponding to the CMB simulation of Fig.~\ref{fig:map_temp} 
(which has been normalised to unit dispersion, as mentioned before).
 Note that regions 
with positive values of $\kappa_{G}$ correspond to hill or lake points
while negative values correspond to saddle points. This behaviour is 
similar to the one found for the determinant, but for $\kappa_{G}$ the gradient
 weights this information in those regions where the later takes large values.    
    
\subsubsection{The extrinsic curvature}
The extrinsic curvature is defined by equation (\ref{eq:extrinsic_curvature}).
To obtain an analytical expression for the pdf of this scalar is very 
complicated and, therefore, we do not include it here. 

%However, we show 
%the average histogram of the extrinsic curvature obtained from 20 CMB 
%normalised Gaussian simulations (bottom panel of Fig.~\ref{curvatures}). 

\begin{figure}
\begin{center}
\includegraphics[width=\hsize]{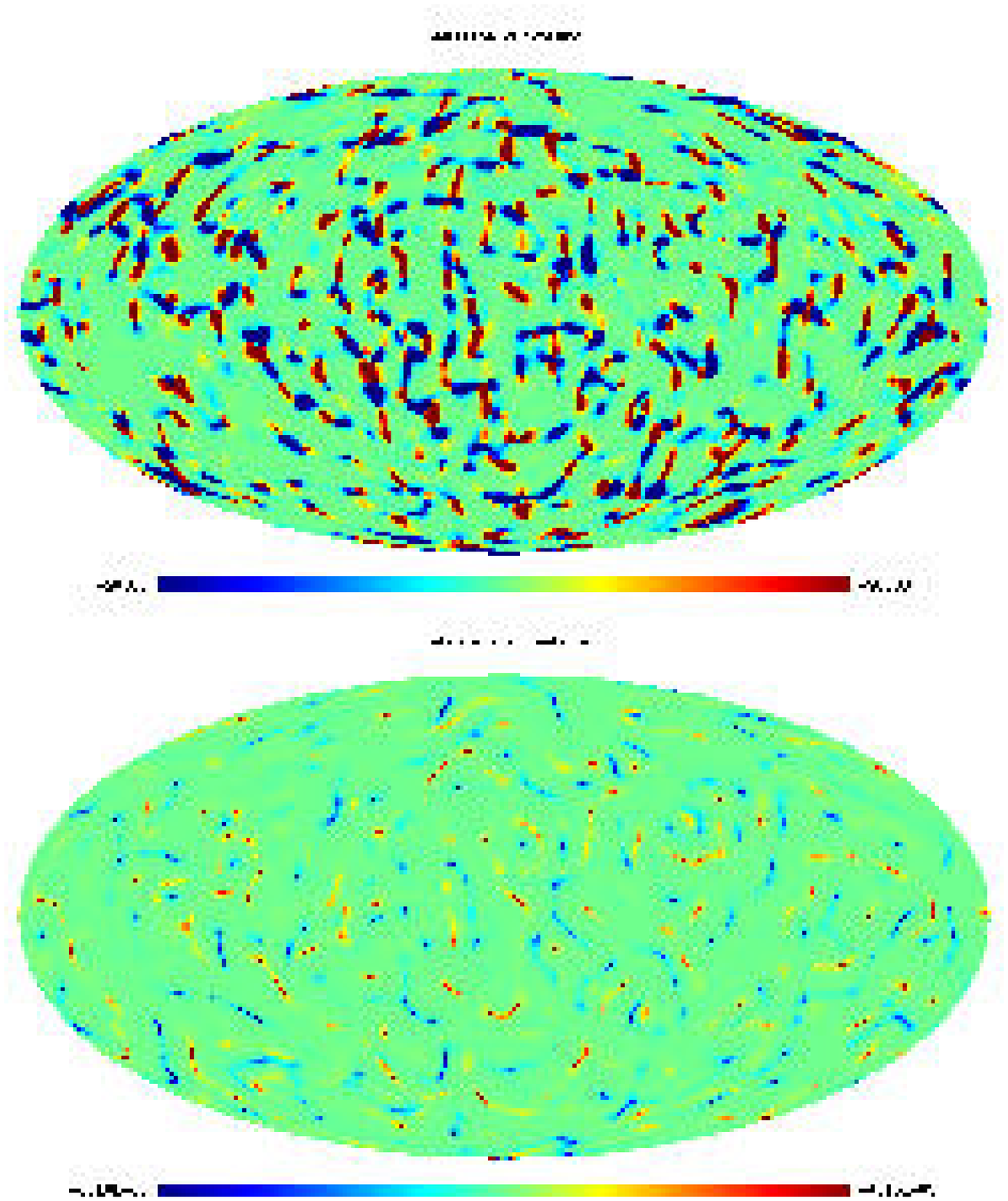}
\caption{The maps of curvature scalars corresponding to the Gaussian
CMB simulation given in Fig.~\ref{fig:map_temp}
normalised to unity dispersion, (so we work with the adimensional map 
$\Delta T / \sigma_{0}$.), are shown: Gaussian curvature (top) and extrinsic 
curvature (bottom)}
\label{fig:map_curv}
\end{center}
\end{figure}

However, we show in the bottom panel of Fig.~\ref{fig:map_curv} the extrinsic 
curvature map corresponding to the normalised CMB simulation of 
Fig.~\ref{fig:map_temp}.
%Significado fisico
%Tambien hay q cambiar esta figura
%

\section{Normalised Scalars}

In order to enhance possible non-Gaussian signatures in the temperature field, 
it would be desirable to study quantities which are independent of the power 
spectrum of the analysed field. With this aim, we have constructed a new set 
of scalars that have this property and that are related to the physical 
scalars of the previous sections. In addition, as will be shown later, these 
new quantities are mathematically simpler, and allows one to deal in a 
straightforward manner with anisotropic fields.  

In some cases the relation between the normalised and the ordinary scalar, 
is only a constant factor. In this case the distribution function of these 
new quantities, can be deduced straightforwardly from the old ones. We can 
include in this group the square modulus of the gradient and its derivative, 
the Laplacian, the distortion, the shear and the ellipticity. The rest of the 
new scalars should be defined using other normalised scalars, and their pdf's 
constructed accordingly (see Apendix C). 

In the next subsection, we introduce the normalised scalars and give their 
distribution function for the Gaussian case. We have summarized this 
information in table ~\ref{tab:norm_scalars}.
\begin{table*}
\begin{center}
    \begin{tabular}{ c  c  c  c  c }
      \hline
       Normalised scalar & Notation &  Definition & Domain&  pdf \\
      \hline
greatest eigenvalue & $\tilde{\lambda}_{1}$& $\frac{1}{2} \left( \tilde{\lambda}_{+}+ \tilde{\lambda}_{-} \right)$ & $(\tilde{\lambda}_{2},\infty)$ & $p(\tilde{\lambda}_{1})= \frac{4}{3\sqrt{2\pi}} e^{-2\tilde{\lambda}_{1}^{2}} \left( 1 + \sqrt{\frac{2\pi}{3}} \tilde{\lambda}_{1} e^{2 \frac{\tilde{\lambda}_{1}^{2}}{3}} \left[ 1 + erf\left(\sqrt{\frac{2}{3}} \tilde{\lambda}_{1}\right) \right] \right)$   \\
      \hline
lowest eigenvalue & $\tilde{\lambda}_{2}$& $\frac{1}{2} \left( \tilde{\lambda}_{+} - \tilde{\lambda}_{-} \right)$ & $(-\infty,\tilde{\lambda}_{1})$ & $p(\tilde{\lambda}_{2})= \frac{4}{3\sqrt{2\pi}} e^{-2\tilde{\lambda}_{1}^{2}} \left( 1 - \sqrt{\frac{2\pi}{3}} \tilde{\lambda}_{1} e^{2 \frac{\tilde{\lambda}_{1}^{2}}{3}} \left[ 1 - erf\left(\sqrt{\frac{2}{3}} \tilde{\lambda}_{1}\right) \right] \right)$ \\
      \hline
Laplacian & $\tilde{\lambda}_{+}$& $-\frac{\lambda_{+}}{\sigma_{2}}$ & $(-\infty,\infty)$ & $p(\tilde{\lambda}_{+})=\frac{1}{\sqrt{2\pi}} e^{\frac{-\tilde{\lambda}_{+}^{2}}{2}}$  \\
      \hline
determinant & $\tilde{d}$ & $\tilde{\lambda}_{1} \tilde{\lambda}_{2}$ & $(-\infty,\infty)$ & $p(\tilde{d}) = \left\{ \begin{array}{ll} \frac{4}{\sqrt{3}} e^{4\tilde{d}} & \tilde{d} < 0 \\
\frac{4}{\sqrt{3}} e^{4\tilde{d}} \left[1 - erf\left( \sqrt{6\tilde{d}}\right)    \right] &  \tilde{d} > 0 \\
\end{array} \right .$   \\
      \hline
shear & $\tilde{y}$ & $\frac{y}{\sigma_{2}^{2}-2\sigma_{1}^{2}}$ & $(0,\infty)$ & $p(\tilde{y})=4e^{-4\tilde{y}}$ \\
      \hline
distortion & $\tilde{\lambda}_{-}$& $\frac{\lambda_{-}}{\sqrt{\sigma_{2}^{2}-2\sigma_{1}^{2}}}$ & $(0,\infty)$ & $p(\tilde{\lambda}_{-})=2 \tilde{\lambda}_{-} e^{-\tilde{\lambda}_{-}^{2}}$               \\
      \hline
ellipticity & $\tilde{e}$ & $\frac{e\sigma_{2}}{\sqrt{\sigma_{2}^{2}-2\sigma_{1}^{2}}}$ & $(-\infty,\infty)$ &  $p(\tilde{e})=4 \vert \tilde{e}\vert \left( 1+8\tilde{e}^2 \right)^{-\frac{3}{2}}$                             \\ 
      \hline 
shape index & $\tilde{\iota}$& $\frac{2}{\pi} \arctan{\left(-\frac{1}{2\tilde{e}}\right)}$ & $(-1,1)$ & $p(\tilde{\iota})= \frac{\pi}{2} \frac{\vert \cos{\left(\frac{\pi}{2} \tilde{\iota} \right)}\vert }{\vert \sin^{3}{\left(\frac{\pi}{2} \tilde{\iota}\right)} \vert} \left[ 1 + 2 \cot^{2}{\left(\frac{\pi}{2} \tilde{\iota}\right)}\right]^{-\frac{3}{2}}$                      \\
      \hline
gradient & $\tilde{g}$& $\frac{g}{\sigma_{1}^{2}}$ & $(0,\infty)$ & $p(\tilde{g})=e^{-\tilde{g}}$ \\
      \hline
derivative of gradient & $\tilde{D}_{g}$& $\frac{D_{g}}{\sqrt{8} \sigma_{r} \sigma_{p}^{2}}$ & $(-\infty,\infty)$ & $p(\tilde{D}_{g})=\frac{2}{\sqrt{\pi}} \int_{0}^{\infty}{e^{-y^{2}-\frac{\tilde{D}_{g}^2}{y^{4}}} \frac{\it{d}y}{y}}$ \\
      \hline 
Gaussian curvature & $\tilde{\kappa}_{G}$ & $\frac{\tilde{d}}{\left( 1 + \tilde{g}\right)^{2}} $ & $(-\infty,\infty)$  & $p(\tilde{\kappa}_{G}) = \left\{ \begin{array}{ll}
-\int_{-\infty}^{\tilde{\kappa}_{G}}\frac{2e}{\tilde{\kappa}_{G}} \sqrt{\frac{z}{3\tilde{\kappa}_{G}}}  e^{-\sqrt{\frac{z}{\tilde{\kappa}_{G}}}} e^{4z} dz & \tilde{\kappa}_{G} < 0 \\
\int_{\tilde{\kappa}_{G}}^{\infty}\frac{2e}{\tilde{\kappa}_{G}} \sqrt{\frac{z}{3\tilde{\kappa}_{G}}}  e^{-\sqrt{\frac{z}{\tilde{\kappa}_{G}}}} e^{4z} \left[ 1 -  erf \left( \sqrt{6z}\right)\right] dz & \tilde{\kappa}_{G} > 0 \\
\end{array}
\right .$       \\
      \hline 
    \end{tabular}
    \caption{\label{tab:norm_scalars}List of the normalised scalars. The pdf's 
are valid for any homogeneous and isotropic Gaussian random field, 
independently of its power spectrum.}
    \end{center}
  \end{table*}

In order to test the theoretical distributions obtained for the normalised 
scalars for the Gaussian case, we have generated 20 full-sky CMB Gaussian 
simulations, in the way explained in section~\ref{gaussian_case}. A
map of the studied normalised scalar is then obtained for each simulation 
and a distribution function of the scalar is constructed. Finally, the
average value and dispersion of the distribution function from the 20
simulations is computed and plotted versus the theoretical distribution.

\subsection{The normalised Hessian matrix scalars}

\subsubsection{The normalised eigenvalues}

\begin{figure*}
\includegraphics[angle=0, width=14.0cm]{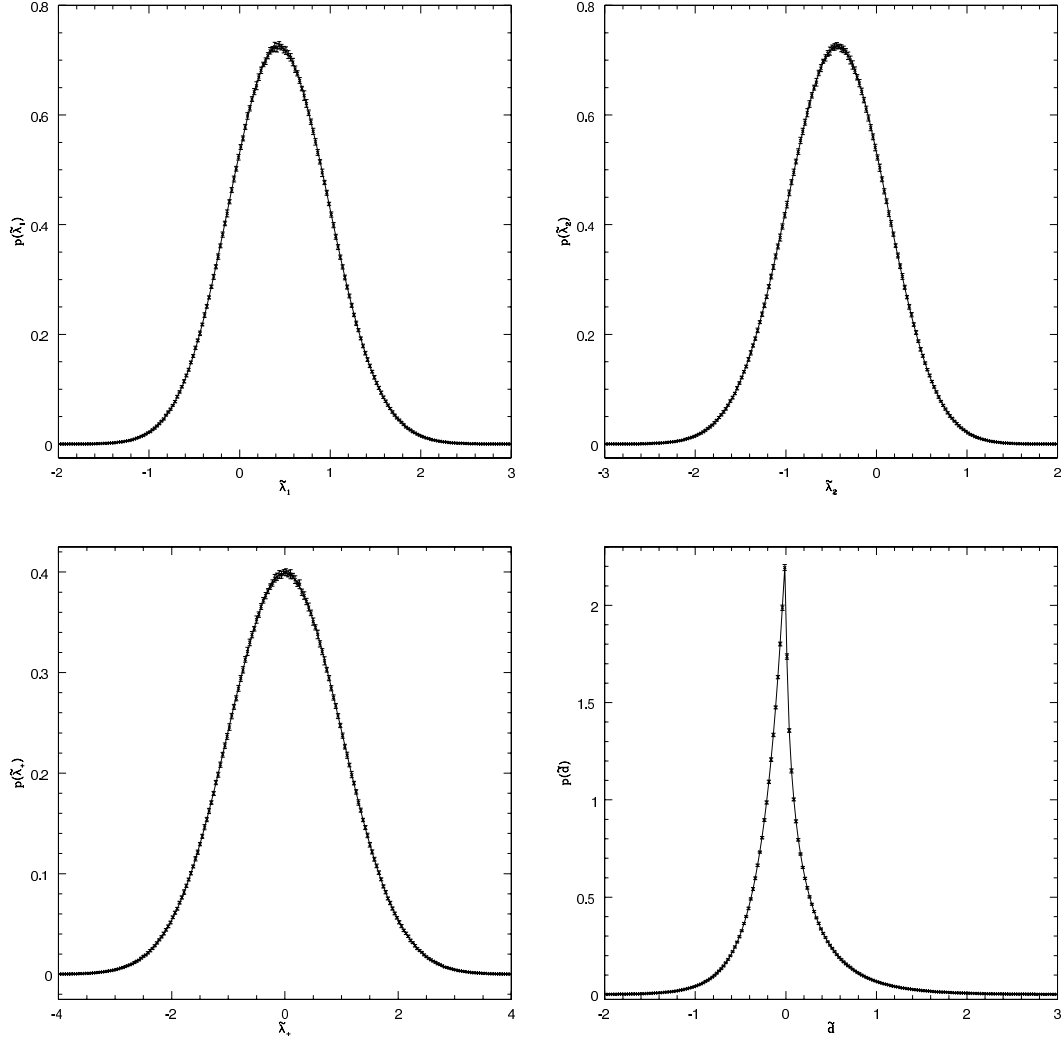}
%\includegraphics[angle=0,
%width=7.0cm]{./IMAGES/hist_elipticity_3ns_200bines.ps} 
%\includegraphics[angle=0, width=7.0cm]{./IMAGES/hist_shape_index_3ns.ps}
%\includegraphics[angle=0, width=7.0cm]{./IMAGES/hist_lambda_1_3ns.ps}
%\includegraphics[angle=0, width=7.0cm]{./IMAGES/hist_lambda_2_3ns.ps}
\caption{Theoretical distributions and results obtained from
CMB Gaussian simulations for the normalised eigenvalues (top left and right), 
the normalised Laplacian (bottom left) and the normalised determinant 
(bottom right). The normalised Laplacian is Gaussian distributed whereas the 
other three normalised scalars follow different non-Gaussian distribution.}
\label{fig:local_curvatures_univ}
\end{figure*}

The normalised eigenvalues are defined in terms of the original eigenvalues 
through the following expression:
\begin{equation}
\left( \begin{array}{c}
\tilde{\lambda}_{1}  \\
\tilde{\lambda}_{2} \end{array} \right) = \frac{1}{2}
\left( \begin{array}{cc}
\frac{1}{\sigma_{2}} + \frac{1}{\sqrt{\sigma_{2}^{2}-2\sigma_{1}^{2}}}  &  \frac{1}{\sigma_{2}} - \frac{1}{\sqrt{\sigma_{2}^{2}-2\sigma_{1}^{2}}} \\
\frac{1}{\sigma_{2}} - \frac{1}{\sqrt{\sigma_{2}^{2}-2\sigma_{1}^{2}}}  &  \frac{1}{\sigma_{2}} + \frac{1}{\sqrt{\sigma_{2}^{2}-2\sigma_{1}^{2}}}  \end{array} \right)
\left( \begin{array}{c}
\lambda_{1}  \\
\lambda_{2} \end{array} \right)
\label{eq:matriz_cv_eigen}
\end{equation}
Therefore analogously to the original eigenvalues, 
$\tilde{\lambda}_{2} < \tilde{\lambda}_{1} < \infty$ and 
$-\infty < \tilde{\lambda}_{2} < \tilde{\lambda}_{1}$.
Note that for a power spectrum as the one of the CMB, we have 
$\sigma_{1}<<\sigma_{2}$. Using this approximation we find for the previous 
equation that the normalised eigenvalues are approximately proportional to 
the original ones and, therefore, have basically the same physical meaning. 
Their distribution functions for the Gaussian case are given by (see
Appendix \ref{ap:norm_scalars}): 
\begin{equation}
p(\tilde{\lambda}_{1})= \frac{4
e^{-2\tilde{\lambda}_{1}^{2}}}{3\sqrt{2\pi}} \left\{ 1 +
\sqrt{\frac{2\pi}{3}} \tilde{\lambda}_{1} e^{2
\frac{\tilde{\lambda}_{1}^{2}}{3}} \left[ 1 + {\rm
erf}\left(\sqrt{\frac{2}{3}} \tilde{\lambda}_{1}\right) \right] \right\}  
\end{equation}
\begin{equation}
p(\tilde{\lambda}_{2})= \frac{4
e^{-2\tilde{\lambda}_{2}^{2}}}{3\sqrt{2\pi}} \left\{ 1 -
\sqrt{\frac{2\pi}{3}} \tilde{\lambda}_{2} e^{2
\frac{\tilde{\lambda}_{2}^{2}}{3}} \left[ 1 - {\rm
erf}\left(\sqrt{\frac{2}{3}} \tilde{\lambda}_{2}\right) \right] \right\}
\end{equation}
The pdf for $\tilde{\lambda}_1$ and $\tilde{\lambda}_2$, together with the 
corresponding results obtained from the simulations, are shown in the top 
panels of Fig.~\ref{fig:local_curvatures_univ}.

\subsubsection{The normalised Laplacian}

The normalised Laplacian, $\tilde{\lambda}_{+}$, is defined using the 
original Laplacian:
\begin{equation}
\tilde{\lambda}_{+}=-\frac{\lambda_{+}}{\sigma_{2}}
\end{equation}
where, by definition $-\infty < \tilde{\lambda}_{+} < \infty$. The normalised 
Laplacian of a HIGRF follows a Gaussian distribution with zero mean and unit 
dispersion: 
\begin{equation}
p(\tilde{\lambda}_{+})=\frac{1}{\sqrt{2\pi}} e^{\frac{-\tilde{\lambda}_{+}^{2}}{2}}.
\label{eq:pdf_normalised_lambda_+}
\end{equation}
 Fig.~\ref{fig:local_curvatures_univ} (bottom left panel) shows that the
distribution obtained for $\tilde{\lambda}_{+}$ from the Gaussian CMB 
simulations perfectly follows its theoretical pdf.

\subsubsection{The normalised determinant}

The normalised determinant, $\tilde{d}$, is defined analogously to the 
original one, using the normalised scalars $\tilde{\lambda}_{1}$ and 
$\tilde{\lambda}_{2}$ as follows:
\begin{equation}
\tilde{d} = \tilde{\lambda}_{1} \tilde{\lambda}_{2}
\end{equation}
where $\tilde{d} \in (-\infty,\infty)$. Using the approximation 
$\sigma_{1}<< \sigma_{2}$ valid for CMB maps, the normalised determinant 
becomes proportional to the original determinant.  

The probability density function of 
the normalised determinant for the Gaussian case is (See Appendix 
~\ref{ap:norm_scalars}).
\begin{equation}
p(\tilde{d}) = \left\{ \begin{array}{ll}
\frac{4}{\sqrt{3}} e^{4\tilde{d}} & \tilde{d} < 0 \\
\frac{4}{\sqrt{3}} e^{4\tilde{d}} \left[1 - erf\left( \sqrt{6\tilde{d}}\right)    \right] &  \tilde{d} > 0 \\
\end{array}
\right .
\label{eq:pdf_normalised_det_A}
\end{equation}
The bottom right panel of Fig.~\ref{fig:local_curvatures_univ} shows
that the agreement between the theoretical pdf and that obtained from 
simulated CMB maps is very good.

\subsection{The normalised distortion scalars}

\subsubsection{The normalised shear}

The normalised shear, $\tilde{y}$, is defined as a constant factor times the 
original shear,   
\begin{equation}
\tilde{y}= \frac{y}{\sigma_{2}^{2}- 2 \sigma_{1}^{2}}
\end{equation}
so inherited by the original shear, $0< \tilde{y} < \infty $. The distribution 
function of this normalised quantity for the Gaussian case is simpler than 
for $y$ :
\begin{equation}
p(\tilde{y})= 4 e^{- 4 \tilde{y}}.
\end{equation}
The comparison between this probability density function and the one obtained 
from 20 Gaussian CMB simulations, is shown 
in the top left panel of Fig.~\ref{fig:shear,ellipticity & shape index_univ}. 
Note that the agreement between the theoretical curve and the results obtained 
from simulations is very good. 
\begin{figure*}
\includegraphics[angle=0, width=14.0cm]{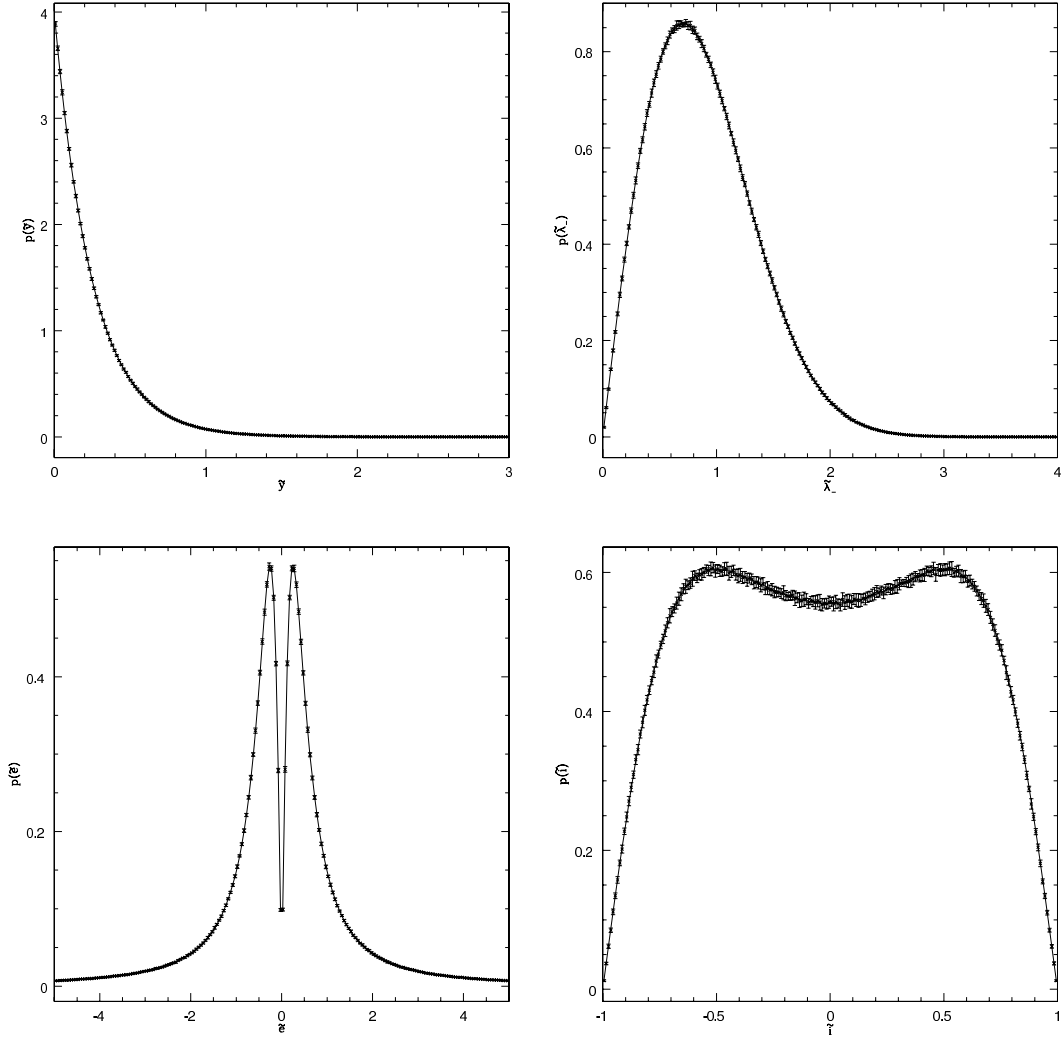}
\caption{Theoretical distributions compared to the results from
CMB Gaussian simulations for the normalised shear (top left), the normalised 
distortion (top right), the normalised ellipticity (bottom left) and the 
normalised shape index (bottom right).}
\label{fig:shear,ellipticity & shape index_univ}
\end{figure*}

\subsubsection{The normalised distortion}

The normalised distortion, $\tilde{\lambda}_{-}$, is defined by:   
\begin{equation}
\tilde{\lambda}_{-} =  \frac{\lambda_{-}}{\sqrt{\sigma_{2}^{2}-2\sigma_{1}^{2}}} 
\label{eq:norm_lambda_{-}}
\end{equation}
and analogously to the original distortion, 
$ 0 < \tilde{\lambda}_{-} < \infty$. The probability density function of the 
normalised distortion for a HIGRF is
\begin{equation} 
p(\tilde{\lambda}_{-})=2 \tilde{\lambda}_{-} e^{-\tilde{\lambda}_{-}^{2}}.
\label{eq:pdf_normalised_lambda_-}
\end{equation}
This theoretical distribution versus the one obtained from simulations are 
given in the top right panel of Fig.\ref{fig:shear,ellipticity & shape index_univ}.

\subsubsection{The normalised ellipticity}
\label{norm_elip}

The normalised ellipticity, $\tilde{e}$, is proportional to $e$ :
\begin{equation}
\tilde{e}=\frac{\sigma_{2}}{\sqrt{\sigma_{2}^{2}-2\sigma_{1}^{2}}} e
\end{equation}
where, by construction $-\infty < \tilde{e} < \infty$. The probability density 
function of the normalised ellipticity of a HIGRF is 
\begin{equation}
p(\tilde{e})=4 \vert \tilde{e}\vert \left( 1+8\tilde{e}^2 \right)^{-\frac{3}{2}}.
\label{eq:norm_ellipticity_pdf}
\end{equation}
Note that this distribution function has very long tails and that
$<\tilde{e}^2>$ is not defined. This problem was inherited from the
original ellipticity pdf, and motivates the introduction 
of the shape index, which is a bounded quantity.

The theoretical pdf and the results obtained from CMB simulations are
compared in the bottom left panel of Fig.~\ref{fig:shear,ellipticity &
shape index_univ}. 

\subsubsection{The normalised shape index}

The normalised shape index, $\tilde{\iota}$, is defined in terms of the normalised 
ellipticity in an analogous way to the definition of the original shape index, 
(see equation (\ref{eq:shape_index_def})):
\begin{equation}
\tilde{\iota}= \frac{2}{\pi} \arctan{\left( -\frac{1}{2\tilde{e}} \right)}
\label{eq:def_normalised_shape_index}
\end{equation}
Therefore $-1 < \tilde{\iota} < 1$. The distribution function of the normalised 
shape index, for the Gaussian case, is deduced from the one of the normalised 
ellipticity through simple transformations:
\begin{equation}
p(\tilde{\iota})= \frac{\pi}{2} \frac{\vert \cos{\left(\frac{\pi}{2} \tilde{\iota} \right)}\vert }{\vert \sin^{3}{ \left( \frac{\pi}{2} \tilde{\iota} \right)} \vert} \left[ 1 + 2 \cot^{2}{\left( \frac{\pi}{2} \tilde{\iota} \right)}\right]^{-\frac{3}{2}}.
\end{equation}
The bottom right panel of Fig.~\ref{fig:shear,ellipticity & shape index_univ} 
shows the good agreement between the theoretical prediction and the results 
obtained from simulations for the normalised shape index.

\subsection{The normalised gradient related scalars}

\subsubsection{The normalised square of the gradient modulus}

The normalised square of the gradient modulus, $\tilde{g}$, is defined as the 
original gradient times a constant factor:
\begin{equation}
\tilde{g}= \frac{g}{\sigma_{1}^{2}},
\end{equation} 
therefore $0 < \tilde{g} < \infty$ and the resulting distribution function of 
this normalised quantity for the Gaussian case, is simply
\begin{equation}
p(\tilde{g})=e^{-\tilde{g}}.
\end{equation}
The top panel of Fig.~\ref{gradient,Dg_univ} shows the theoretical 
distribution function compared to the results obtained from
simulations for $\tilde{g}$, showing excellent agreement.
\begin{figure}
\begin{center}
\includegraphics[angle=0, width=7.0cm]{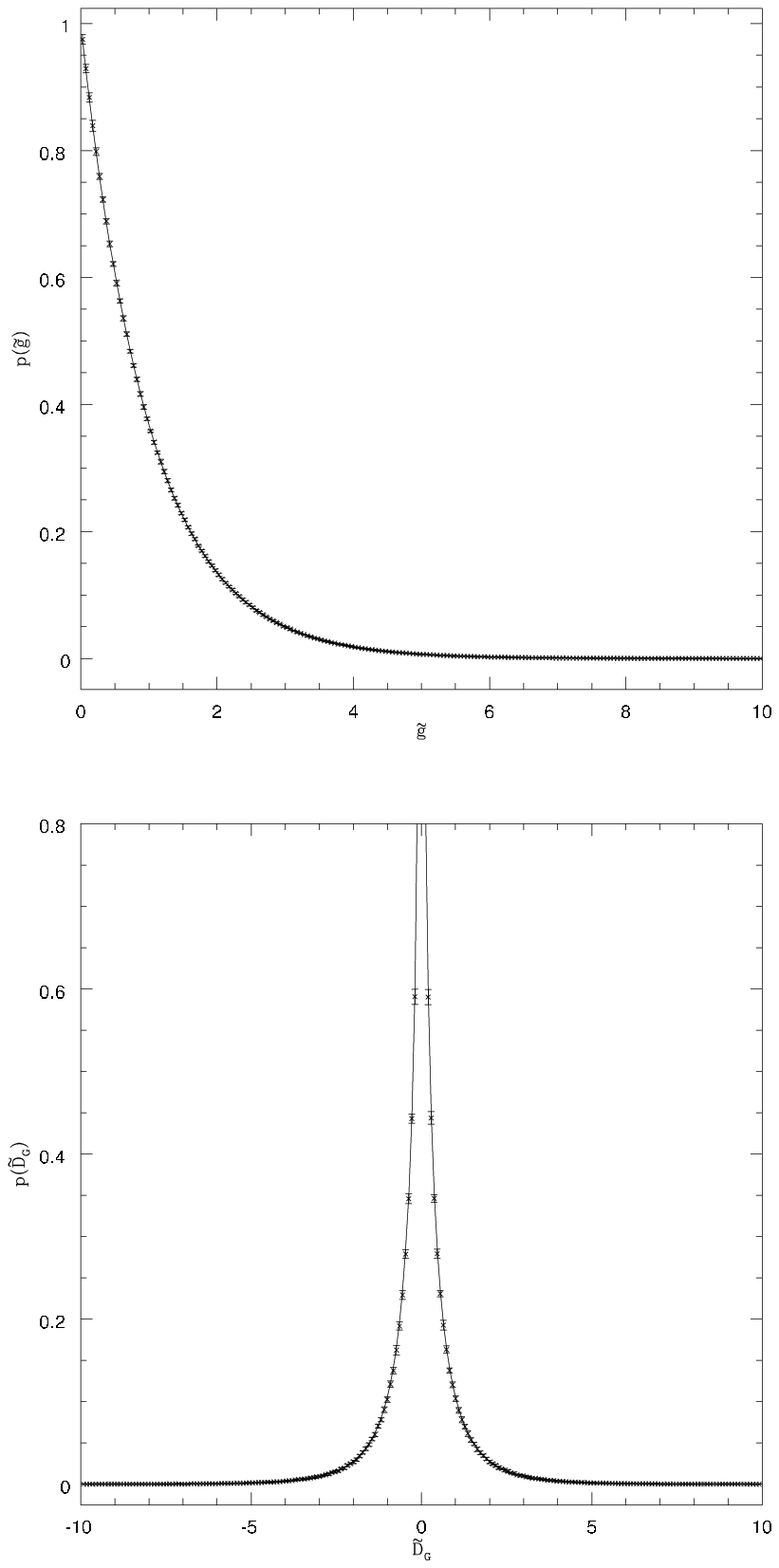}
%\includegraphics[angle=0, width=7.0cm]{./IMAGES/hist_gradcuad_3ns.ps}
%\includegraphics[angle=0, width=7.0cm]{./IMAGES/hist_omega_3_3ns.ps}
%width=7.0cm]{./IMAGES/hist_gaussian_curvature_3ns.ps} 
%\includegraphics[angle=0, width=7.0cm]{./IMAGES/hist_omega_3_3ns.ps}
\caption{Comparison between the theoretical pdf and the one
obtained from Gaussian CMB simulations for the normalised square of the 
gradient modulus (top) and the normalised derivative of the gradient.}
\label{gradient,Dg_univ}
\end{center}
\end{figure}

\subsubsection{The normalised derivative of the square of gradient modulus}

The normalised derivative of the square of the gradient modulus, 
$\tilde{D_{g}}$, is proportional to the original scalar:
\begin{equation}
\tilde{D}_{g}= \frac{D_{g}}{\sqrt{8} \sigma_{r} \sigma_{p}^{2}}, 
\end{equation}
so inherited by the original scalar, $-\infty < \tilde{D}_{g} < \infty$. The 
probability density function of this normalised scalar for the Gaussian case, 
is trivially obtained from the pdf of the primitive scalar:
\begin{equation}   
p(\tilde{D}_{g})=\frac{2}{\sqrt{\pi}} \int_{0}^{\infty}{e^{-y^{2}-\frac{\tilde{D}_{g}^2}{y^{4}}} \frac{\it{d}y}{y}}.
\end{equation}
The bottom panel of Fig.~\ref{gradient,Dg_univ} shows the theoretical pdf and 
the one obtained from simulations for $\tilde{D}_{g}$. Again the agreement is very good.

\subsection{The normalised curvature scalars}

\subsubsection{The normalised Gaussian curvature}

\begin{figure}
\begin{center}
\includegraphics[angle=0, width=7.0cm]{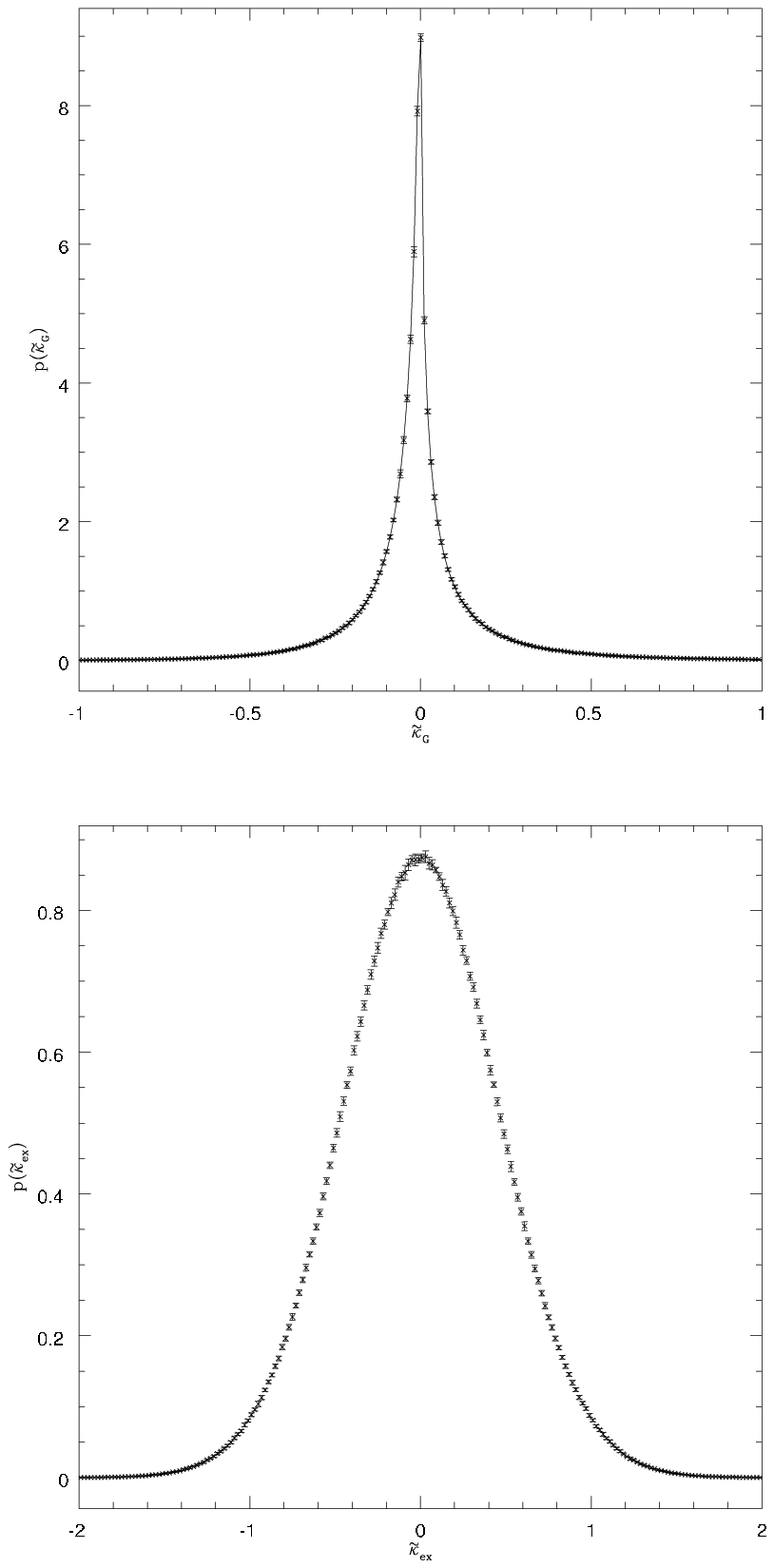}
\caption{Comparison between the theoretical pdf and the one
obtained from 20 Gaussian CMB simulations for the normalised Gaussian curvature (top), and distribution function obtained from simulations for the normalised extrinsic curvature (bottom).}
\label{curvatures_univ}
\end{center}
\end{figure}

The normalised Gaussian curvature, $\tilde{\kappa}_{G}$, is defined in terms of the 
normalised determinant and the normalised gradient, as follows:
\begin{equation}
\tilde{\kappa}_{G} = \frac{\tilde{d}}{\left( 1 + \tilde{g}\right)^{2}}, 
\end{equation}
where by construction $-\infty < \tilde{\kappa}_{G} < \infty $.
% This normalised quantity 
%is defined as the division of two uncorrelated quantities. 
The pdf of the normalised Gaussian curvature of a HIGRF is given by: 
\begin{equation}
p(\tilde{\kappa}_{G}) = \left\{ \begin{array}{ll}
-\int_{-\infty}^{\tilde{\kappa}_{G}}\frac{2e^{4z+1}}{\tilde{\kappa}_{G}} \sqrt{\frac{z}{3\tilde{\kappa}_{G}}}  e^{-\sqrt{\frac{z}{\tilde{\kappa}_{G}}}} dz & \tilde{\kappa}_{G} < 0 \\
\int_{\tilde{\kappa}_{G}}^{\infty}\frac{2e^{4z+1}}{\tilde{\kappa}_{G}} \sqrt{\frac{z}{3\tilde{\kappa}_{G}}} e^{-\sqrt{\frac{z}{\tilde{\kappa}_{G}}}} \\ 
~ ~ \left[ 1 -  erf \left( \sqrt{6z}\right)\right] dz & \tilde{\kappa}_{G} > 0 \\
\end{array}
\right .
\end{equation}
The comparison between the theoretical and simulated results are shown
in the top panel of Fig.~\ref{curvatures_univ}. 

\subsubsection{The normalised extrinsic curvature}

The normalised extrinsic curvature, $\tilde{\kappa}_{ex}$, is defined 
analogously to the original scalar, using equation 
(\ref{eq:extrinsic_curvature}):
\begin{equation}
\tilde{\kappa}_{ex}=\frac{1}{2} ~ \frac{1}{\sqrt{1+\tilde{g}}}\left[\tilde{\lambda}_{+} -
\frac{\tilde{D}_{g}}{\left(1+\tilde{g}\right)}\right] ~,
\label{eq:normalised_extrinsic_curvature}
\end{equation}
Therefore $-\infty < \tilde{\kappa}_{ex} < \infty$. As for the
original ${\kappa}_{ex}$, it is very complicated to obtain an
analytical expression for the pdf of this normalised scalar.
However we have  included in the bottom panel of Fig
\ref{curvatures_univ}, the pdf of the  
normalised extrinsic curvature obtained from 20 Gaussian CMB simulations. 

\subsection{Correlations between normalised scalars}
By looking at the definitions of the different scalars, it becomes
apparent that some of the scalars are related to each other.
In order to know how much independent information they contain, it is
interesting to study the correlations between the different scalars.
Using simulations, we have calculated these correlations for a
HIGRF which are given in table \ref{tab:correlations}. 
We have included all the normalised scalars except for the
ellipticity, due to the fact that the 
dispersion of this quantity is not defined and, therefore, the usual
correlation coefficient can not be calculated. However, the ellipticity and the
shape index are very closely related and thus it is expected that the
correlations between the ellipticity and the rest of the scalars will
be similar to those obtained for the shape index.

It is interesting to note that the gradient of the modulus, $\tilde{g}$, is 
the only scalar which is uncorrelated with all the considered scalars. On the 
contrary, the eigenvalues, $\tilde{\lambda}_{1}$ and $\tilde{\lambda}_{2}$, 
have some degree of correlation (or anticorrelation) with all the scalars 
(except for $\tilde{g}$). Within the Hessian scalars, we see that only the 
normalised Laplacian, $\tilde{\lambda}_{+}$, and determinant, $\tilde{d}$, are 
uncorrelated between them. $\tilde{\lambda}_{+}$ is also uncorrelated with two 
of the normalised distortion scalars (the shear, $\tilde{y}$, and the 
distorsion, $\tilde{\lambda}_{-}$) as well as with Gaussian curvature 
$\tilde{\kappa}_{g}$. Conversely, $\tilde{d}$ has some correlation with these 
three scalars but is uncorrelated with the normalised shape index, $\tilde{\iota}$,
 the derivative of the gradient, $\tilde{D}_{g}$ and extrinsic curvature, 
$\tilde{\kappa}_{ex}$. With regard to the distortion scalars, $\tilde{y}$ and 
$\tilde{\lambda}_{-}$ are strongly correlated, which is expected 
since $\tilde{y}$ is, except by a constant factor, the square of
$\tilde{\lambda}_{-}$. However, these two scalars are uncorrelated with
the shape index $\tilde{\iota}$. It is also interesting to note that 
$\tilde{\kappa}_{G}$ is uncorrelated with $\tilde{\kappa}_{ex}$.

%The first order Gaussian variables p and q, of section \ref{gaussian_case}, 
%are uncorrelated of second order ones: r, s and t. Therefore the normalised 
%gradient must be uncorrelated with second order normalised scalars (Hessian 
%matrix and distorsion scalars). All these second order scalars
%containing the information of two of them (in example
%$\tilde{\lambda}_{1}$ and %$\tilde{\lambda}_{2}$), consequently they
%are correlated. 

%As the normalised scalars are independent of the power spectrum, we
%have chosen Gaussian CMB simulations of $N_{side}=256$ smoothed with a
%FWHM=33arcmin gaussian beam to perform the scalars. 

\begin{table*}
\begin{center}
    \begin{tabular}{c |  c  c  c  c  c  c  c  c  c  c  c }
          & $\tilde{\lambda}_{1}$ & $\tilde{\lambda}_{2}$ &  $\tilde{\lambda}_{+}$ &  $\tilde{d}$  & $\tilde{y}$ &  $\tilde{\lambda}_{-}$ & $\tilde{\iota}$  &  $\tilde{g}$ & $\tilde{D}_{g}$ & $\tilde{\kappa}_{G}$ &  $\tilde{\kappa}_{ex}$  \\
      \hline
      $\tilde{\lambda}_{1}$  & 1.00 & 0.64 & 0.91 & -0.23 & 0.40 & 0.42 & -0.83 & 0 & -0.52 & -0.19 & 0.89 \\
      
      $\tilde{\lambda}_{2}$  & 0.64 & 1.00 & 0.91 & 0.23 & -0.40 & -0.42 & -0.83 & 0 & -0.52 & 0.19 & 0.89 \\

      $\tilde{\lambda}_{+}$  & 0.91 & 0.91 & 1.00 & 0 & 0 & 0 & -0.92 & 0 & -0.58 & 0 & 0.99 \\

      $\tilde{d}$  & -0.23 & 0.23 & 0 & 1.00 & -0.58 & -0.55 & 0 & 0 & 0 & 0.83 &  0 \\

      $\tilde{y}$  & 0.40 & -0.40 & 0 & -0.58 & 1.00 & 0.96 & 0 & 0 & 0 & -0.48 & 0 \\

      $\tilde{\lambda}_{-}$  & 0.42 & -0.42 & 0 & -0.55 & 0.96 & 1.00 & 0 & 0 & 0 & -0.46 & 0 \\

%      $\tilde{e}$  & & & & & & & & & & &  \\

      $\tilde{\iota}$  & -0.83 & -0.83 & -0.92 & 0 & 0 & 0 & 1.00 & 0 & 0.53 & 0 & -0.91 \\

      $\tilde{g}$  & 0 & 0 & 0 & 0 & 0 & 0 & 0 & 1.00 & 0 & 0 & 0 \\

      $\tilde{D}_{g}$  & -0.52 & -0.52 & -0.58 & 0 & 0 & 0 & 0.53 & 0 & 1.00 & 0 & -0.57 \\

      $\tilde{\kappa}_{G}$  & -0.19 & 0.19 & 0 & 0.83 & -0.48 & -0.46 & 0 & 0 & 0 & 1.00 &  0 \\
      $\tilde{\kappa}_{ex}$  & 0.89 & 0.89 & 0.99 & 0 & 0 & 0 & -0.91 & 0 & -0.57 & 0 & 1.00 \\
    \end{tabular}
    \caption{\label{tab:correlations}
Correlations between the normalised scalars. }
    \end{center}
  \end{table*}

\section{Treatment of instrumental noise and masks}

Real data contain not only the cosmological signal but also contaminating 
emissions and instrumental noise. The Galactic region, where the Galactic 
foregrounds dominate, is usually masked from the data. It is also a common 
procedure to mask the emission coming from extragalactic point sources. 
These masked pixels are then discarded from the analysis.

In addition, the instrumental noise produces discontinuities from pixel to 
pixel in the map, since this noise is assumed to be, in general, uncorrelated 
(white noise). The discontinuities introduce problems in the calculation of 
the derivatives. In order to deal with this issue, we will smooth the CMB 
signal plus noise with a Gaussian beam of FWHM equal to 2.4 times the pixel 
size (where we assume that this is the size of the Gaussian beam used to 
filter the original CMB signal). 
 
Assuming that the white noise is homogeneous and isotropic, its power spectrum 
is given by: 
\begin{equation}
C_{\ell}^{n} = \frac{4 \pi \sigma_{n}^{2}}{N_{pixtot}}
\end{equation}  
where $\sigma_{n}$ is the noise dispersion and $N_{pixtot}$ is the total 
number of pixels. So the resulting power spectrum of the noisy simulations 
can be expressed as follows:  
\begin{equation}
C_{\ell} = [C_{\ell}^{s} e^{-\ell \left(\ell +1 \right) \sigma_{g}^{2}} + C_{\ell}^{n}] e^{-\ell \left(\ell +1 \right) \sigma_{g}^{2}}
\end{equation}
where $\sigma_{g}$ is the Gaussian beam dispersion and $C_{\ell}^{s}$ 
is the CMB power spectrum.

However, in most experiments, the instrumental noise is usually
an anisotropic Gaussian random field, charactarised
by a different dispersion at each pixel $\sigma_n(\vec{x})$ (with
$\vec{x}$ the unitary vector on the sphere in 
the direction of observation). In order to deal with anisotropic
noise, we have built a ``pixel-dependent power spectrum''
$H_{\ell}(\vec{x})$ in the following way:
\begin{equation}
H_{\ell}(\vec{x}) = \left[C_{\ell}^{s} e^{-\ell \left(\ell +1 \right)
\sigma_{g}^{2}} + \frac{4 \pi
\sigma_{n}^{2}(\vec{x})}{N_{pixtot}}\right] e^{-\ell \left(\ell +1
\right) \sigma_{g}^{2}}. 
\label{eq:spectrum_aninoise}
\end{equation}
$H_{\ell}(\vec{x})$ would be the power spectrum of a map that contain
the filtered CMB plus isotropic noise with dispersion
$\sigma_n(\vec{x})$, and filtered again with a Gaussian beam of
disperson $\sigma_{g}$.
In practice, due to this second smoothing, the dispersion of the
noise in a given pixel would depend not only on the noise level on
that position but also of its neighbours. However, provided that the
dispersion of 
the noise varies smoothly, which is usually the case, the previous
equation is a good approximation to the ``power spectrum in each pixel''.

Since, by construction, the normalised scalars are independent of the
power spectrum of the field, we can obtain these quantities {\it for each
pixel}, taking into account equation (\ref{eq:spectrum_aninoise}). We
just need to construct the moment fields on the sphere
$\sigma_{0}(\vec{x}), \sigma_{1}(\vec{x})$  and 
$\sigma_{2}(\vec{x})$ using equation (\ref{eq:momentos}), where the
$C_{\ell}$'s are now given by $H_{\ell}(\vec{x})$. The normalised
scalars are then 
calculated by introducing these pixel-dependent moments on their
corresponding definitions. The probability distribution function
obtained from the map of the normalised scalar constructed in this way
will follow the corresponding theoretical distribution given in the
previous section.
This is one of the advantages of working with the normalised
scalars. Note that if we construct the ordinary scalars for a map
containing anisotropic noise, the resulting probability distribution
will not longer follow the theoretical pdf for a Gaussian field.

In addition to noise we also have a masked region, which introduces 
discontinuities in the boundary of the mask. Moreover, we have no information 
about the field inside these masked regions. As before, in order to deal with 
this problem, the first step is to smooth the masked map (with the pixels of 
the mask set to zero) with a Gaussian beam of FWHM equal to 2.4 times the 
pixel size. As already mentioned, this solves the problem of the 
discontinuities of the noise and also reduces the mask boundary problem. 
However a more sophisticated procedure is necessary to eliminate the effect of 
the mask, since if we calculate the normalised scalars from this smoothed 
masked map, those pixels close to the boundary would be strongly contaminated 
by the spurious signal of the mask. Nevertheless, the values of the normalised 
scalars in those pixels far enough from the mask would be correct. Therefore 
we need to generate an extended mask that eliminates from the analysis not 
only the original masked pixels but also those pixels in the neighborhood of 
the mask. The particular shape of this extended mask would depend on the 
original mask but also on the particular normalised scalar that we are 
considering.

We can obtain the extended mask for each normalised scalars using simulations. 
In particular, we construct and compare one $\it{exact}$ and one 
$\it{approximated}$ map for the considered quantity for each simulated map. 
The exact map $I_{e}$ is constructed, simply, by calculating the normalised 
scalar from the smoothed noisy map; since the derivatives are obtained without
 masking any region, they will be properly calculated. The approximated map 
$I_{a}$ is obtained by calculating the corresponding normalised scalar from 
the masked smoothed noisy map. In this case the derivatives of those pixels 
close to the boundary of the masked region would be heavily contaminated by the 
mask. For each simulation, at each pixel outside the original mask, we 
calculate then the error quantity $\epsilon$ defined as:
\begin{equation}     
\epsilon = \frac{\vert I_{e} - I_{a} \vert}{\sigma_{e}}
\end{equation}
where $\sigma_{e}$ is the dispersion of the exact map for the corresponding 
simulation. The average of this quantity over a large number of simulations 
is then obtained and we have a map of $\epsilon$ at each pixel outside the original 
mask. The extended mask will be formed, in addition to the pixels of the 
original mask, by those pixels with $\epsilon$ greater than a fixed value $\epsilon_{*}$, 
that is, we keep for the analysis only those pixels where the value of the 
considered normalised scalar is reasonably close to the correct value and 
therefore are not appreciably contaminated by the presence of the mask. 
Considering only the pixels outside the extended mask, the
distribution of the  
normalised scalars should follow the theoretical pdf's found in previous 
sections.    

In order to test the method using realistic simulations, we have studied the 
normalised scalars using masked CMB maps containing anisotropic
noise. We have  
used the Kp0 mask, defined in Bennett et al. 2003b, which covers 
approximately a 24 per cent of the sky. 
The anisotropic noise has been simulated with a level equal to that
expected for the noise-weighted average of the
Q, V and W WMAP channels (Komatsu et al. 2003) after two years of
observation. The signal-to-noise ratio per pixel of the simulations
ranges from 1.6 to 5.9 (at resolution $N_{side}=256$).
To obtain the extended mask we have used 100 simulations and
find that a value of $\epsilon_{*} = 0.1$ provides good results.
We have then obtained from the simulations the pdf for each of the
normalised scalars (but using only those pixels outside the extended
mask) and compare it with the theoretical pdf.
Table~\ref{tab:mask_aninoise} gives the percentage of useful pixels
that are kept after applying the corresponding extended mask for the 
normalised scalars.
\begin{table}
\begin{center}
    \begin{tabular}{| c | c |}
      \hline
      Scalar &  Useful pixels (\%) \\
      \hline
      $\tilde{\lambda}_{1}$ & 70.6 \\
      \hline
      $\tilde{\lambda}_{2}$ & 70.5 \\
      \hline
      $\tilde{\lambda}_{+}$  & 70.1 \\
      \hline
      $\tilde{d}$  & 72.0 \\
      \hline
      $\tilde{y}$  & 67.8 \\
      \hline
      $\tilde{\lambda}_{-}$  & 69.2 \\
%      \hline
%      $\tilde{e}$  &    \\
      \hline
      $\tilde{\iota}$  & 68.1 \\      
      \hline
      $\tilde{g}$  & 70.7 \\
      \hline
      $\tilde{D}_{g}$  & 71.8  \\
      \hline
      $\tilde{\kappa}_{G}$  & 70.7 \\
      \hline
      $\tilde{\kappa}_{ex}$ & 70.1 \\ 
      \hline
    \end{tabular}
    \caption{\label{tab:mask_aninoise}
The percentage of useful pixels kept after applying the extended masks
obtained with $\epsilon_{*}=0.1$ for the normalised scalars.}
    \end{center}
  \end{table}
For all the normalised scalars we see that approximately a 70 per cent of the 
pixels can be used for the analysis, loosing only around a 6 per cent of the 
sky with respect to the Kp0 mask. The extended
mask for the normalised ellipticity can not be obtained using the
error quantity $\epsilon$, since the dispersion of this scalar is not defined.
Alternative error quantities could be used to obtain its extended
mask. However, for simplicity, we have used the mask obtained for the
normalised shape index, since these two quantities are closely related.

The pdf's obtained from 100 simulations (using only those pixels outside
the extended mask) for the normalised Laplacian, distortion and shape
index are compared to their corresponding theoretical pdf's showing an
excellent agreement. Similar results are obtained for the rest of
normalised scalars (including the ellipticity), using their corresponding 
extended masks (which are not shown for the sake of brevity).
\begin{figure}
\includegraphics[angle=0, width=7.0cm]{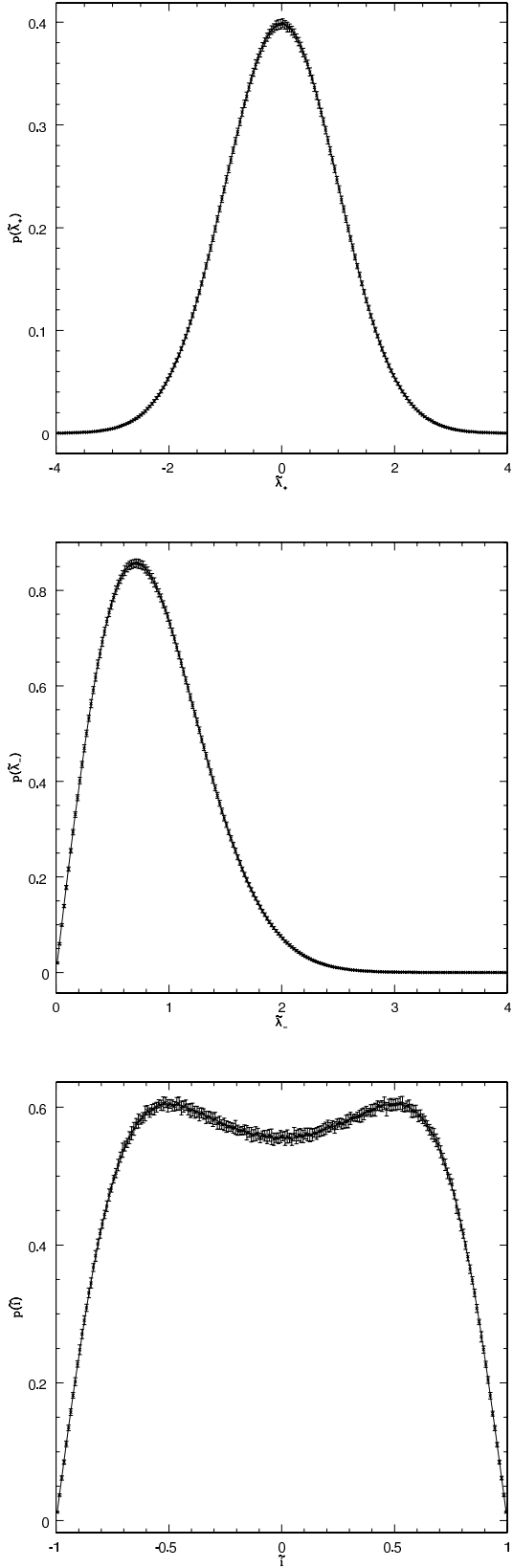}
%\includegraphics[angle=0, width=7.0cm]{./IMAGES/hist_laplac_3ns_noise2.0.ps}
%\includegraphics[angle=0,
%width=7.0cm]{./IMAGES/hist_shape_index_3ns_noise4.0.ps} 
\caption{The distribution obtained from 100 simulations compared to the
corresponding theoretical pdf is shown for three different 
normalised scalars: Laplacian (top), distortion 
(middle) and shape index (bottom). The distributions have been
obtained using only those pixels outside the extended mask (with
$\epsilon_{*}$=0.10).}
\label{fig:mask+noise}
\end{figure}
%

%We would like to point out that for more irregular masks (for instance
%with holes due to extragalactic point sources) the amount of pixels
%lost for the analysis would increase faster than in the simple
%case considered here. Therefore it may be necessary to relax the
%condition $e_*=0.05$ in order to have more useful pixels. For instance we
%have tested that $e_*=0.1$ also leads to good results, but a detailed
%study is recommended in each case.

\section{Conclusions}
In this paper we have introduced several scalar quantities on the
sphere and we have shown how to calculate them for a CMB map directly
from the $a_{\ell m}$ harmonic coefficients. In particular, we have
considered the square modulus of the gradient, the Laplacian, the
determinant of the negative Hessian matrix, the distorsion, the shear, the
ellipticity, the shape index, the eigenvalues of the negative Hessian
matrix, the Gaussian curvature, the extrinsic curvature and the gradient 
derivative. 
Assuming a homogeneous and isotropic Gaussian field, we have derived
analytical or semi-analytical expressions for the pdf's of each of the
scalars. For convenience, we define normalised scalars which have more simple 
distribution functions than ordinary scalars. Therefore, for these 
normalised scalars we have tested the theoretical results with simulations, 
showing an excellent agreement.

We propose these scalars to study the Gaussian character of the CMB. 
In particular, we aim to test their power using non-Gaussian CMB simulations 
based in the Edgeworth expansion (Mart\'\i nez-Gonz\'alez et al. 2002). 
Moreover, we plan to apply this method in future works to both physically 
motivated non-Gaussian models and the WMAP data. Finally, we would like 
to point out that the study of
the scalars can be adapted, for instance, to deal with regions or
extrema above (or below) different thresholds. This will also be
studied in a future work.

\section*{Acknowledgments}
The authors thank Patricio Vielva for useful comments.
CM thanks the Spanish Ministerio de Ciencia y Tecnolog\'{\i}a (MCyT) for a
predoctoral FPI fellowship. RBB thanks UC and the MCyT for a Ram{\'o}n
y Cajal contract. We acknowledge partial financial support from the
Spanish MCyT project ESP2004- 07067-C03-01.
This work has used the software package HEALPix (Hierarchical, Equal
Area and iso-latitude pixelization of the sphere,
http://www.eso.org/science/healpix), developed by K.M. G\'orski,
E.F. Hivon, B.D. Wandelt, A.J. Banday, F.K. Hansen and
M. Barthelmann.
We acknowledge the use of the software package CMBFAST
(http://www.cmbfast.org) developed by U. Seljak and M. Zaldarriaga.

\appendix
\section{Derivatives on the sphere}
\label{ap:derivatives}

In this appendix we derive the harmonic coefficients of the first and
second derivatives of the temperature field as a function of the
$a_{\ell m}$. With these expressions and using the HEALPix package it
is straightforward to calculate the maps of the field derivatives as
well as the scalars.
  
It is a common procedure to write the temperature field on the sphere
as a series of harmonic functions:
\begin{equation}
T(\theta,\phi) = \sum_{\ell ,m} a_{\ell m} Y_{\ell m}(\theta,\phi) ~.
\label{eq:expansion}
\end{equation}
The spherical harmonic functions constitute an orthonormal complete
set. Also, they are eigenvectors of the momentum operators
$L_z$ and $L^2$:
\begin{eqnarray}
L_{z} Y_{\ell m} &=& m Y_{\ell m} ~,\\
L^2 Y_{\ell m} &=& \ell (\ell +1)Y_{\ell m} ~.  
\end{eqnarray}
In addition, the $L_{+}$ and $L_{-}$ operators that are defined:
\begin{equation}
L_{\pm} = L_{x} \pm i L_{y} ~,
\end{equation}
act on the spherical harmonic functions in the following way:
\begin{equation}
L_{\pm} Y_{\ell m}=\sqrt{\left(\ell \pm m + 1\right)\left(\ell \mp
m\right)} Y_{\ell m\pm1}~.
\end{equation}
In order to obtain the derivatives of the field, we will write them as
a function of the previous operators:
\begin{eqnarray}
\frac{\partial}{\partial\theta}&=&\frac{1}{2}\left[e^{-i\phi}
L_{+}-e^{i\phi}L_{-}\right] ~, 
\label{eq:derivatives1}
\end{eqnarray}
\begin{eqnarray} 
\frac {\partial}{\partial\phi} &=&i L_{z} ~.
\label{eq:derivatives2}
\end{eqnarray}
On the other hand, we will also make use of the following common
spherical harmonic relations:
\begin{eqnarray}
e^{-i\phi}\sin{\theta} ~ Y_{\ell m+1}&=&\sqrt{\frac{\left(\ell
-m\right)\left(\ell -m+1\right)}{\left(2\ell +1\right)\left(2\ell
+3\right)}}Y_{\ell +1m}- \nonumber \\  
&-&\sqrt{\frac{\left(\ell +m\right)\left(\ell +m+1\right)}{\left(2\ell
-1\right)\left(2\ell +1\right)}}Y_{\ell -1 m} ~, 
\end{eqnarray}
\begin{eqnarray} 
e^{i\phi}\sin{\theta} ~ Y_{\ell m-1}&=&\sqrt{\frac{\left(\ell
+m\right)\left(\ell +m+1\right)}{\left(2\ell +1\right)\left(2\ell
+3\right)}}Y_{\ell +1m}- \nonumber \\ 
&-&\sqrt{\frac{\left(\ell -m\right)\left(\ell -m+1\right)}{\left(2\ell
-1\right)\left(2\ell +1\right)}} Y_{\ell -1 m}  ~,
\end{eqnarray}
\begin{eqnarray}
\cos{\theta} ~ Y_{\ell
m}&=&\sqrt{\frac{\left(\ell-m+1\right)\left(\ell+m+1\right)}{\left(2\ell+1\right)\left(2\ell+3\right)}}Y_{\ell+1
m} + \nonumber \\ 
&+&
\sqrt{\frac{\left(\ell+m\right)\left(\ell-m\right)}{\left(2\ell+1\right)\left(2\ell-1\right)}} 
Y_{\ell-1 m} ~. 
\end{eqnarray}

Applying the derivative operators given in (\ref{eq:derivatives1}) and 
(\ref{eq:derivatives2}) to
the temperature field (\ref{eq:expansion}), and taking into account
the previous spherical harmonic relations, it is straightforward to
obtain the harmonic coefficients of the fields $\frac{\partial T}
{\partial \phi}$ and $\sin \theta \frac{\partial T}{\partial \theta}$,
$b_{\ell m}$ and  $c_{\ell m}$ respectively, as a function of the harmonic
coefficients $a_{\ell m}$ of the original field\footnote{Note that
similar expressions to obtain the first derivatives of the
temperature field have been independently obtained by Eriksen et
al. (2004b). We also want to remark that in Schmalzing \& Gorski 1998 a 
similar method to obtain the derivatives on the sphere was described.}
\begin{equation}
c_{\ell m}=im ~ a_{\ell m} ~,
\end{equation}
\begin{equation}
b_{\ell m}=H_{+}\left(\ell,m\right) a_{\ell+1,m} + H_{-}\left(\ell,m\right) a_{\ell-1,m} ~,
\end{equation}
where $H_{\pm}$ are functions of $\ell,m$ given by
\begin{equation}
H_{+}\left(\ell,m\right)=-\left(\ell+2\right)\sqrt{\frac{\left(\ell+m+1\right)\left(\ell-m+1\right)}{\left(2\ell+1\right)\left(2\ell+3\right)}}~,
\end{equation}
\begin{equation}
H_{-}\left(\ell,m\right)=\left(\ell-1\right)\sqrt{\frac{\left(\ell+m\right)\left(\ell-m\right)}{\left(2\ell+1\right)\left(2\ell-1\right)}} ~,
\end{equation} 
Note that these functions are even on $m$ and that $H_-(\ell,\ell)$ is zero
for all the $\ell$ values. 

Analogously, we can obtain the harmonic
coefficients $d_{\ell m}$, $e_{\ell m}$, and $f_{\ell m}$
corresponding to the fields $\frac{\partial^{2} T}{\partial \phi^{2}}$,
$\sin^{2}{\theta} \frac{\partial^{2} T}{\partial \theta^{2}}$ and
$\sin{\theta}\frac{\partial^{2} T}{\partial \theta \partial \phi}$:
\begin{eqnarray}
d_{\ell m}&=&-m^{2}a_{\ell m} ~,
\end{eqnarray}
\begin{eqnarray}
e_{\ell m}&=& \left[\left(\frac{\ell\left(\ell+3\right)}{\left(\ell+2\right)^2}\right)H_{+}^{2}\left(\ell,m\right) \nonumber \right.\\
&-& \left.\left(\frac{\left(\ell+1\right)\left(\ell-2\right)}{\left(\ell-1\right)^{2}}\right)H_{-}^{2}\left(\ell,m\right)\right]a_{\ell m} +  \nonumber \\  
 &+&\left[\left(\frac{\ell+3}{\ell+2}\right)H_{+}\left(\ell,m\right)H_{+}\left(\ell+1,m\right)\right]a_{\ell+2,m} + \nonumber\\ 
 &+& \left[\left(\frac{\ell-2}{\ell-1}\right)H_{-}\left(\ell,m\right)H_{-}\left(\ell-1,m\right)\right]a_{\ell-2,m} ~,
\end{eqnarray}
\begin{eqnarray}
f_{\ell m} &=& im \left[H_{+}\left(\ell,m\right)a_{\ell+1,m}+H_{-}\left(\ell,m\right)a_{\ell-1,m}\right] ~.
\end{eqnarray}

Making use of the previous results and taking into account the
different expressions presented along this work, it is straightforward
to obtain the maps of the considered scalars.

\section{Some useful results regarding spherical harmonic series} 
\label{ap:sums} 
In this appendix we present some useful results for spherical harmonic
series, which are necessary to calculate the covariances of the
variables defined in equations (\ref{eq:xtheta}-\ref{eq:x3}). 

In particular, we have made use of the following expressions which
involve the harmonic functions and their derivatives with respect to
$\theta$: 
\begin{equation}
\sum_{m=-\ell}^{\ell} \frac{\partial Y_{\ell m}}{\partial\theta}
Y_{\ell m}^{*} = 0 ~,
\end{equation}
\begin{equation}
\sum_{m=-\ell}^{\ell} \frac{\partial Y_{\ell
m}}{\partial\theta}\frac{\partial Y_{\ell m}^{*}}{\partial\theta}
=-\sum_{m=-\ell}^{\ell} \frac{\partial^2 Y_{\ell m}}{\partial\theta^2}
Y_{\ell m}^{*} = \frac{2l+1}{4\pi}\frac{l(l+1)}{2} ~, 
\end{equation}
\begin{equation}
\sum_{m=-\ell}^{\ell} \frac{\partial Y_{\ell m}}{\partial\theta}\frac{\partial^2 Y_{\ell m}^{*}}{\partial\theta^2} = 0 ~.
\end{equation}

Another interesting set of series involve the presence of powers of
$m$. In particular, we need:
\begin{equation}
\sum_{m=-\ell}^{\ell} m^4 Y_{\ell m}^{2} = \frac{2\ell+1}{4\pi} \left[\frac{3}{8}\left[\ell\left(\ell+1\right)\right]^2 - \frac{1}{2} \ell\left(\ell+1\right) \left[\frac{\cos{2\theta}}{\sin^{2}{\theta}}\right]\right] ~,
\end{equation}
\begin{equation}
\sum_{m=-\ell}^{\ell} m^2  Y_{\ell m}^2(\theta,\phi) = \frac{2\ell+1}{4\pi}\frac{l(l+1)}{2} \sin^2{\theta} ~, 
\end{equation}
and applying $\frac{\partial}{\partial\theta}$ to the previous
expression, we get:
\begin{equation}
\sum_{m=-\ell}^{\ell} m^2 \frac{\partial Y_{\ell m}}{\partial\theta} Y_{\ell m}^{*} = \frac{2l+1}{4\pi}\frac{l(l+1)}{2} \sin{\theta}\cos{\theta} ~.
\end{equation}

It can also be be shown that other series that appear in the
calculation of the covariances of the variables
(\ref{eq:xtheta}-\ref{eq:x3}) and that involve odd powers of $m$ are
zero, such as:

\begin{equation}
\sum_{m=-\ell}^{\ell} m^i Y_{\ell m}^{2} = 0 ~,
\end{equation}

\begin{equation}
\sum_{m=-\ell}^{\ell} m^i Y_{\ell m} \frac{\partial Y_{\ell
m}^{*}}{\partial\theta} = 0 ~,
\end{equation}

\begin{equation}
\sum_{m=-\ell}^{\ell} m^i  \left[\frac{\partial Y_{\ell m}}{\partial
\theta}\right]^{2} = 0 ~,
\end{equation}

\begin{equation}
\sum_{m=-\ell}^{\ell} m^i Y_{\ell m} \frac{\partial^{2} Y_{\ell
m}^{*}}{\partial \theta^{2}} = 0 ~, 
\end{equation}
where $i$ is an odd integer.

\section{Normalised scalars distribution functions}
\label{ap:norm_scalars} 

In this appendix we give some guidelines on how to deduce the pdf of ordinary 
scalars for a HIGRF. We also show how to obtain the probability 
density function of the normalised scalars from that of the original
scalars. As an illustration, we focus on the normalised 
ellipticity and determinant, but similar calculations are needed to
obtain the pdf's of the rest of the scalars.

As explained in section ~\ref{derivatives_scalars}, ordinary scalars can be 
constructed in terms of the Gaussian variables $\{p,q,r,s,t\}$ given in 
equations (\ref{eq:xtheta}) to (\ref{eq:x3}).
In particular, the ellipticity can be rewritten as: 
\begin{equation}
e = -\frac{\sqrt{( r - s )^2 + (2t)^2}}{2( r + s )} ~.
\end{equation}
Let us define the uncorrelated Gaussian variables $R = r + s$, $S = r - s$ and
$T = 2 t$. It is straightforward to show that $\sigma_{R}^{2}=\sigma_{2}^{2}$ 
and $\sigma_{S}^{2} = \sigma_{T}^{2} = \frac{1}{2} \sigma_{2}^{2} - 
\sigma_{1}^{2}$. In terms of these new variables, the ellipticity is given by
\begin{equation}
e = -\frac{\sqrt{S^2+T^2}}{2R} ~.
\end{equation}
Since S, T and R are independent variables, the numerator and the 
denominator are also independent.

The numerator is the square root of the addition of the square of two Gaussian
uncorrelated variables with the same dispersion $\sigma_{S}$. Therefore the 
numerator, $z$, should follow a Rayleigh distribution: 
\begin{equation}
p(z)=\frac{1}{\sigma^2_{S}} y e^{-\frac{z^2}{2 \sigma^2_{S}}},
\end{equation}
where $z>0$. The denominator is Gaussian distributed with zero mean and 
$2\sigma_{R}$ dispersion. Performing a change of variable we can obtain
the pdf of the inverse of the denominator $x \equiv \frac{1}{2R}$:
\begin{equation}
p(x)=\frac{1}{x^2} \frac{1}{2 \sqrt{2 \pi}\sigma_{R}} e^{-\frac{1}{8
x^2 \sigma^2_{R}}} 
\end{equation}
where $-\infty < x < \infty$. Taking into account that $x$ and $z$ are
independent, we have: 
\begin{equation}
p(x,z)= p(x)p(z)= \frac{1}{2\sqrt{2 \pi}\sigma_{R}\sigma^2_{S}} \frac{z}{x^2}
e^{-\frac{z^2}{2 \sigma^2_{S}}} e^{-\frac{1}{8 x^2 \sigma^2_{R}}}
\end{equation}
Since we need to calculate the pdf of $e = x z$, we perform a change
of variables  $(x , z) \to (e , z)$ and then, integrating over z (note that
$z>0$), we obtain the distribution function of the ellipticity given
in equation (\ref{eq:pdf_ellipticity}). 
                                                    
The normalised ellipticity is proportional to the ordinary ellipticity, 
$\tilde{e}=c e$, and, therefore, the pdf of $e$ can be trivially
obtained from $p(e)$ by a simple change of variables: 
\begin{equation}
p(\tilde{e})=\frac{1}{c} p \left(e=\frac{\tilde{e}}{c} \right)               
\end{equation}
The preceding expresion leads to equation (\ref{eq:norm_ellipticity_pdf}). 
The same applies for all the normalised scalars which are proportional to 
the ordinary ones. 

However, some normalised scalars are related to the ordinary ones in a more
complex way. To construct their pdf's it is convenient, whenever possible,
to rewrite them in terms of $\tilde{\lambda}_{+}$ and
$\tilde{\lambda}_{-}$, since these two 
normalised scalars are independent. This can be proved using the
property $p_{\lambda_{1}}(x) = p_{\lambda_{2}}(-x)$, already mentioned
in section ~\ref{gaussian_case}, where $p_{\lambda_{i}}$ denotes the pdf 
of $\lambda_i$. As an example, we show how to calculate the probability 
density function of the normalised determinant, $\tilde{d}$, which can be 
expressed as:
\begin{equation}
\tilde{d}=\tilde{\lambda}_{1} \tilde{\lambda}_{2} = \frac{1}{4} [\tilde{\lambda}_{+}^{2} -\tilde{\lambda}_{-}^{2}]
\end{equation}  
Using the probability distribution function of $\tilde{\lambda}_{+}$ and 
$\tilde{\lambda}_{-}$, (equations~\ref{eq:pdf_normalised_lambda_+})
and~\ref{eq:pdf_normalised_lambda_-}) and by performing different
changes of variables, we can obtain the distribution
function of the normalised determinant. First we calculate the
pdf of the $u\equiv \tilde{\lambda}_{+}^2$ and $v\equiv
\tilde{\lambda}_{-}^2$, which are given by:
\begin{equation}
p(u) =  \frac{1}{\sqrt{2 \pi}} \frac{1}{\sqrt{u}} e^{-\frac{u}{2}}
\end{equation}
\begin{equation}
p(v) =  e^{-v}
\end{equation}
where $0 < u < \infty$ and  $0 < v < \infty$. Since
$\tilde{\lambda}_{+}$ and $\tilde{\lambda}_{-}$  
are independent, $u$ and $v$ are also independent, therefore
\begin{equation}
p(u,v) = p(u) p(v) = \frac{1}{\sqrt{2 \pi}} \frac{1}{\sqrt{u}} e^{-v} e^{-\frac{u}{2}}~.
\end{equation}
We perform then a change of variables 
$(u , v) \to (u , \tilde{d})$, where 
$\tilde{d} = \frac{u}{4}  - \frac{v}{4}$. Finally, integrating over $u$ we
obtain the pdf of the normalised determinant $\tilde{d}$ given in
equation (\ref{eq:pdf_normalised_det_A}).

The normalised eigenvalues can also be rewritten in terms of 
$\tilde{\lambda}_{+}$ and $\tilde{\lambda}_{-}$, through the following
expression:
\begin{equation}
\left( \begin{array}{c}
\tilde{\lambda}_{1}  \\
\tilde{\lambda}_{2} \end{array} \right) =
\left( \begin{array}{cc}
\frac{1}{2} & \frac{1}{2}  \\
\frac{1}{2} & -\frac{1}{2} \end{array} \right)
\left( \begin{array}{c}
\tilde{\lambda}_{+}  \\
\tilde{\lambda}_{-} \end{array} \right),
\label{eq:def_eigenvalues}
\end{equation}
Therefore, throught a straightforward change of variables, 
the pdf of the normalised  eigenvalues can be obtained from those of
$\tilde{\lambda}_{+}$ and $\tilde{\lambda}_{-}$ .

The pdf's of the remaining normalised scalars can be constructed in an
analogous way, since they can be trivially written either in terms of 
$\tilde{\lambda}_{+}$ and $\tilde{\lambda}_{-}$ or of the other 
normalised scalars.

%%%%%%%%%APENDICE C%%%%%%%%%%%%
%$\lambda_{1}$ and $\lambda_{2}$ are correlated variables. We are interested 
%on rewriting the normalised eigenvalues in terms of uncorrelated quantities as 
%$\tilde{\lambda}_{+}$ and $\tilde{\lambda}_{-}$ (See Appendix C).
%%
%\begin{equation}
%\left( \begin{array}{c}
%\tilde{\lambda}_{1}  \\
%\tilde{\lambda}_{2} \end{array} \right) =
%\left( \begin{array}{cc}
%\frac{1}{2} & \frac{1}{2}  \\
%\frac{1}{2} & -\frac{1}{2} \end{array} \right)
%\left( \begin{array}{c}
%\tilde{\lambda}_{+}  \\
%\tilde{\lambda}_{-} \end{array} \right),
%\label{eq:def_eigenvalues}
%\end{equation}
%%
%Supposing a Gaussian initial field, the distribution functions of their 
%eigenvalues are both deduced from the normalised Laplacian and normalised 
%distortion ones, through equation (\ref{eq:def_eigenvalues}).

\end{document}